\documentclass[pra,aps,twocolumn,nopacs,superscriptaddress,nofootinbib,longbibliography]{revtex4}

\usepackage{amsmath}  \usepackage{amssymb}  \usepackage{amsfonts}  \usepackage{bm}  \usepackage{bbm}   \usepackage{braket}  \usepackage{color}  \usepackage{comment}  \usepackage{dcolumn}  \usepackage{enumerate}  \usepackage{epsfig}  \usepackage{graphicx}  \usepackage{indentfirst}  \usepackage{lmodern}  \usepackage{mathrsfs}  \usepackage{mathtools}  \usepackage{psfrag}  \usepackage{pst-all}  \usepackage{soul}  \usepackage{xcolor}  
\usepackage{float} 
\usepackage[colorlinks,linkcolor=blue,citecolor=blue,urlcolor=blue,hyperindex,driverfallback=dvipdfm]{hyperref}  \usepackage[T1]{fontenc} 

\def\ii{{\rm i}}  \def\ee{{\rm e}}
  \def\kB{{k_{\rm B}}}
\def\Ree{{\rm Re}}  \def\Imm{{\rm Im}}
\def\Ab{{\bf A}}        \def\Eb{{\bf E}}  \def\eb{{\bf e}}              \def\Jb{{\bf J}}  \def\jb{{\bf j}}                \def\qb{{\bf q}}  \def\Rb{{\bf R}}  \def\rb{{\bf r}}      \def\vb{{\bf v}} 
\def\xx{\hat{\bf x}}  \def\yy{\hat{\bf y}}  \def\zz{\hat{\bf z}}            
\def\kpar{k_\parallel}  \def\kparb{{\bf k}_\parallel} 
        
\def\Abh{\hat{\bf A}}

\def\Hint{\hat{\mathcal{H}}_{\rm int}} \def\dens{\rho}  \def\aa{\hat{a}}
 \def\Bop{\hat{N_1}} \def\Gop{\hat{N_2}}
\def\Abh{\hat{\bf A}}  
\def\HH{\hat{\mathcal{H}}}  \def\Abh{\hat{\bf A}}  
    \def\Ah{\hat{A}}

\begin{document}
\def\bibsection{\section*{\refname}}

\title{Radiative loss of coherence in free electrons: a long-range quantum phenomenon}

\author{Cruz~I.~Velasco}
\affiliation{ICFO-Institut de Ciencies Fotoniques, The Barcelona Institute of Science and Technology, 08860 Castelldefels (Barcelona), Spain}
\author{Valerio~Di~Giulio}
\affiliation{ICFO-Institut de Ciencies Fotoniques, The Barcelona Institute of Science and Technology, 08860 Castelldefels (Barcelona), Spain}
\author{F.~Javier~Garc\'{\i}a~de~Abajo}
\email{javier.garciadeabajo@nanophotonics.es}
\affiliation{ICFO-Institut de Ciencies Fotoniques, The Barcelona Institute of Science and Technology, 08860 Castelldefels (Barcelona), Spain}
\affiliation{ICREA-Instituci\'o Catalana de Recerca i Estudis Avan\c{c}ats, Passeig Llu\'{\i}s Companys 23, 08010 Barcelona, Spain}

\begin{abstract}{\bf \noindent
Quantum physics rules the dynamics of small objects as they interact over microscopic length scales. Nevertheless, quantum correlations involving macroscopic distances can be observed between entangled photons as well as in atomic gases and matter waves at low temperatures. The long-range nature of the electromagnetic coupling between charged particles and extended objects could also trigger quantum phenomena over large distances. Here, we reveal a manifestation of quantum mechanics that involves macroscopic distances and results in a nearly complete depletion of coherence associated with which-way free-electron interference produced by electron--radiation coupling in the presence of distant extended objects. This is a ubiquitous effect that we illustrate through a rigorous theoretical analysis of a two-path electron beam interacting with a semi-infinite metallic plate and find the inter-path coherence to vanish proportionally to the path separation at zero temperature and exponentially at finite temperature. The investigated regime of large distances involves the coupling of the electron to radiative modes assisted by diffraction at material structures but without any involvement of material excitations. Besides the fundamental interest of this macroscopic quantum phenomenon, our results suggest an approach to measuring the vacuum temperature and nondestructively sensing the presence of distant objects.
}\end{abstract}

\maketitle

\section{INTRODUCTION}

The wave nature of electrons allows us to image materials with atomic resolution in transmission electron microscopy \cite{E05,MKM08} (TEM) and resolve the atomic structure and dynamics of molecules and crystal surfaces through low-energy \cite{DG1927,P1974}, photoemission \cite{paper037}, and ultrafast \cite{WPL16,FML22} electron diffraction. In these techniques, wave interference takes place between elastically scattered components, while inelastic collisions are typically regarded as a source of decoherence that destroys interference through the addition of a stochastic phase.

Decoherence can be produced by coupling to material excitations. In particular, an electron split into two paths and moving parallel to a lossy planar surface was proposed \cite{APZ97}, extensively studied from a theoretical viewpoint \cite{APZ97,MPV03,L04,HL06,M06_3,H11_2,SB12}, and experimentally confirmed \cite{SH07,H10_2,BZB18,KRS20,CB20} to be a suitable configuration to observe electron decoherence. In a related scenario, inelastic electron scattering generated by coupling to thermally populated low-energy material excitations was shown to render an observable loss of electron coherence that limits spatial resolution in TEM \cite{UMH13,UMZ15}. An extension to decoherence of charged particles trapped near a lossy surface has recently been made \cite{MHS22}.

Electron decoherence is equally produced by inelastic excitations associated with photon emission and electromagnetic vacuum fluctuations, as predicted for an electron prepared in a prescribed two-path configuration \cite{F93,F97}, including the effect of neighboring perfect-conductor boundaries \cite{F93,MPV03,HL06}. Likewise, radiative electron decoherence is anticipated to take place due to bremsstrahlung emission \cite{BP01}, interaction with time-varying fields \cite{HF04}, and the Smith-Purcell effect \cite{AM08}. Intriguingly, recoherence can occur for electrons moving in a squeezed vacuum \cite{HF08}.

\begin{figure*}[th!]
\centering{\includegraphics[width=0.8\textwidth]{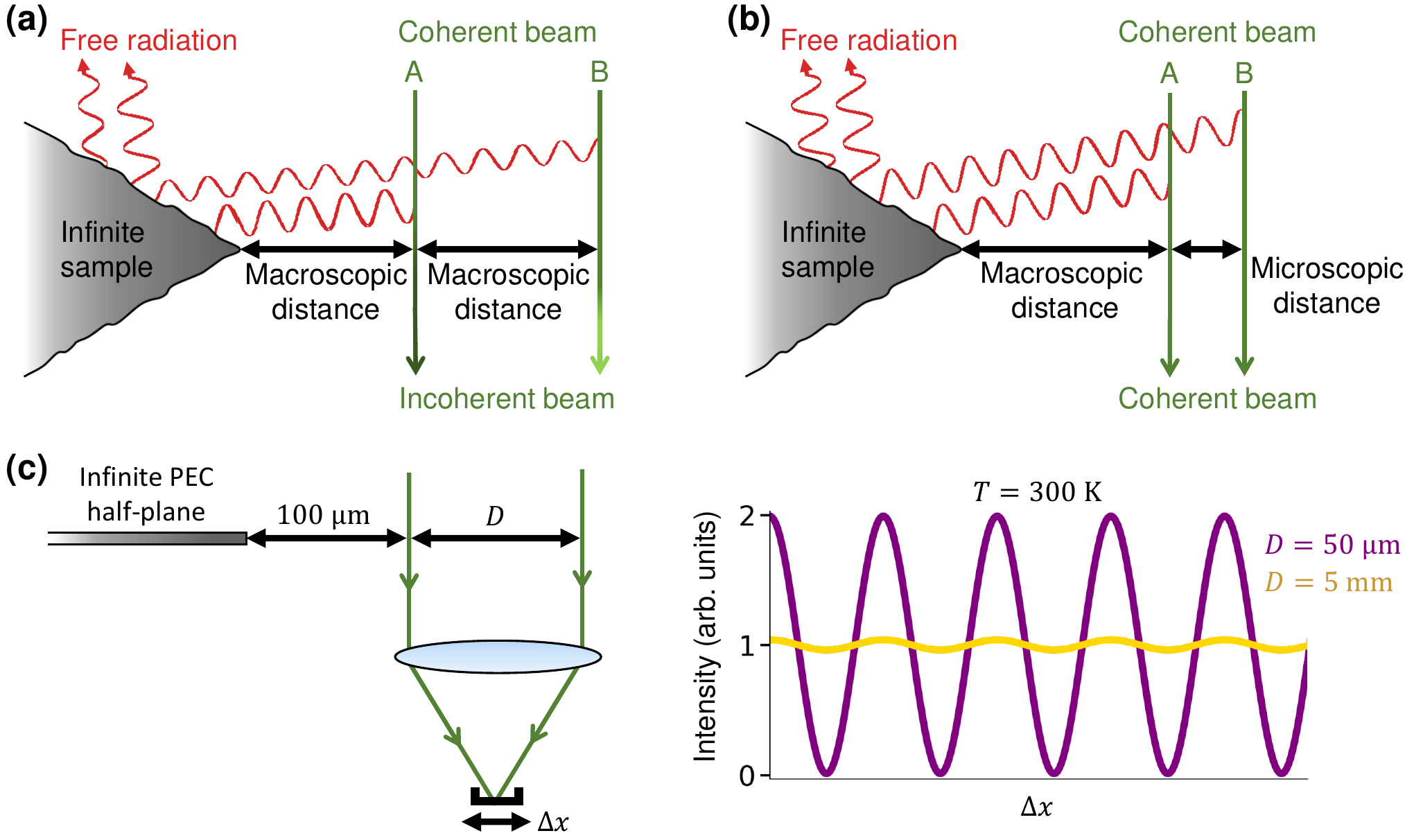}}
\caption{{\bf Electron-beam decoherence due to radiative coupling assisted by extended scatterers.} (a)~An electron split into two paths A and B separated by a large (macroscopic) distance undergoes strong decoherence through coupling to radiation assisted by a distant extended scatterer. (b)~For small (microscopic) inter-path separations, coherence is however preserved. (c)~We show a specific geometry in which one can substantially vary the degree of decoherence by modifying the inter-path separation for a fixed distance to a perfect-electric-conductor (PEC) half-plane at a temperature of 300~K (see also Fig.~S2 in SI), translating into a radical change in the visibility of interference fringes as a function of transverse position $\Delta x$ at an electron detector (see Supplementary Fig.~\ref{FigS1}).}
\label{Fig1}
\end{figure*}

Decoherence can be intuitively understood through the following analysis for an {\it incident} electron whose wave function $\psi^{\rm inc}=\psi_{\rm A}^{\rm inc}+\psi_{\rm B}^{\rm inc}$ is split into two non-overlapping paths A and B (Fig.~\ref{Fig1}a,b). Scattering by a structure produces an overall post-interaction state $\sum_{n}\big(\ket{\psi_{{\rm A},n}}+\ket{\psi_{{\rm B},n}}\big)\otimes\ket{n}$, where $n$ runs over excitations of the involved materials and the radiation field, while $\psi_{{\rm A},n}$ and $\psi_{{\rm B},n}$ denote the electron wave functions in paths A and B resulting after an excitation $n$ is generated. In an interference experiment, electron fringes are formed at a detection plane where the electron paths overlap (Fig.~\ref{Fig1}c). The amplitude of such fringes $\propto\sum_n{\rm Re}\{\langle\psi_{{\rm A},n}|\psi_{{\rm B},n}\rangle\}$ is contributed by wave function components in which the same mode $n$ is excited by both paths. A certain degree of coherence is then preserved if paths A and B can both excite a given mode $n$ with similar amplitudes, just like in a quantum eraser \cite{Z99_2,MKZ16} that produces a loss of which-way information. This is essentially the principle behind inelastic electron holography \cite{LF00,H05,VBS08}, where interference fringes are observed in energy-filtered inelastically scattered electrons (e.g., after they excite a delocalized plasmon that overlaps both electron paths).

In this analysis, the incident electron is prepared in a pure state characterized by a density matrix $\ket{\psi^{\rm inc}}\bra{\psi^{\rm inc}}$, while after the interaction, we have a mixed state with a density matrix $\sum_n\big(\ket{\psi_{{\rm A},n}}+\ket{\psi_{{\rm B},n}}\big)\big(\bra{\psi_{{\rm A},n}}+\bra{\psi_{{\rm B},n}}\big)$ obtained by tracing out material and radiation degrees of freedom, and the resulting loss of visibility in the interference fringes relates to the creation of such inelastic excitations $\ket{n}$. This is conceptually different from diffraction produced after blocking part of the wave function (e.g., in a two-slit experiment), in which the interference fringes are controlled by the shape of the scattering object, but the electron is transmitted in a pure state of wave function $\psi^{\rm tr}$ and density matrix $\ket{\psi^{\rm tr}}\bra{\psi^{\rm tr}}$ without the involvement of any excitations. Interestingly, information theory has been invoked to quantitatively separate actual decoherence from elastic diffraction \cite{SL18}.

In another conceptually different scenario, one can consider electron interactions with classical fields such as those induced by a laser in the context of photon-induced near-field electron microscopy \cite{BFZ09} (PINEM). The electron then follows a coherent evolution \cite{paper151,FES15}, and therefore, it is characterized by a pure state $\ket{\psi_{\rm PINEM}}$, even if the electron experiences energy changes (e.g., sidebands in PINEM) and those changes are path-dependent. Consequently, the electron density matrix $\ket{\psi_{\rm PINEM}}\bra{\psi_{\rm PINEM}}$ remains pure, and different paths can still interfere (e.g., to achieve spatiotemporal electron compression \cite{paper415}).

In the present work, we are interested in the decoherence produced by the creation of inelastic excitations, and more precisely, radiative modes. Consequently, we consider configurations in which the electron paths do not physically intersect any material (Fig.~\ref{Fig1}a,b), and the electron--boundary distances are sufficiently large to neglect inelastic excitations created inside the material (see below). This is conceptually different from previous investigations for an aloof electron moving parallel to a planar interface, which leads to decoherence by generating material excitations \cite{APZ97,MPV03,L04,HL06,M06_3,H11_2,SB12} as well as an elastic phase due to image interactions even for perfect conductors \cite{F93,paper357}.

\begin{table}[th!]
\centering{\includegraphics[width=0.5\textwidth]{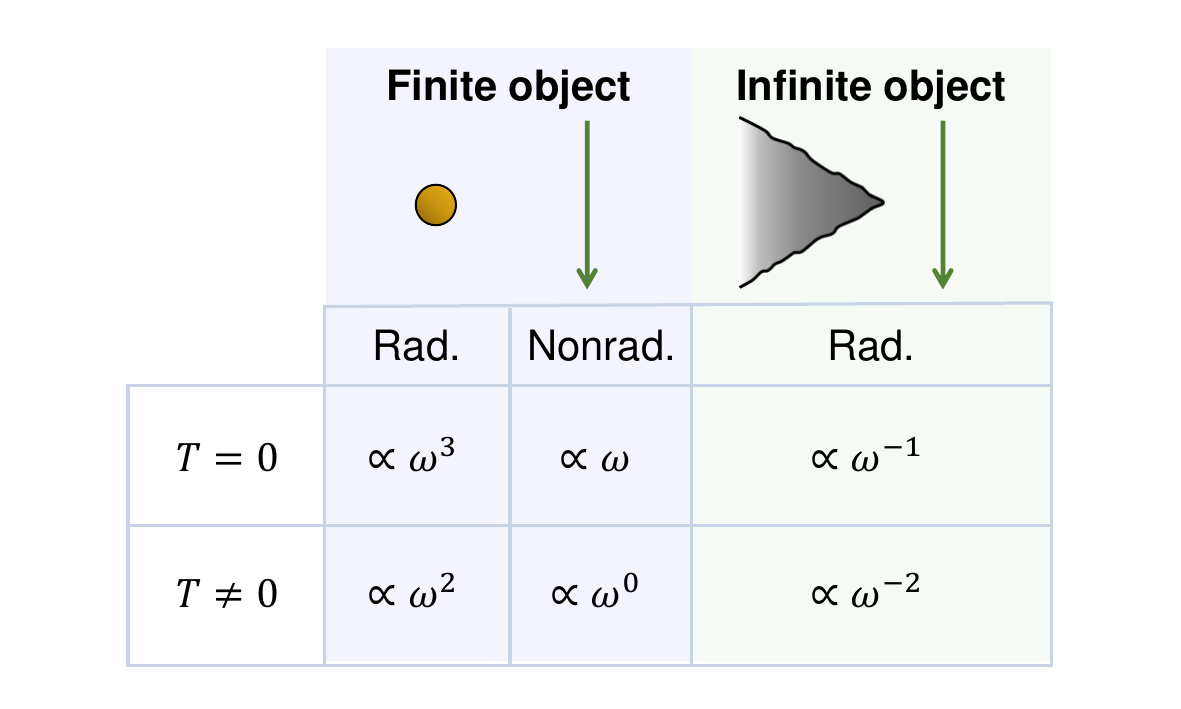}}
\caption{{\bf Divergence in the spectrally resolved electron energy-loss probability.} For finite objects (left), the loss probability $\Gamma(\omega)$ vanishes at low frequencies as they become increasingly small compared to the light wavelength. We have $\Gamma(\omega)\propto\omega^3$ and $\propto\omega$ from radiative and nonradiative losses at zero temperature, while an additional factor of $1/\omega$ appears at finite temperature $T$ because of the scaling of the inelastic-scattering probability as $2n_T(\omega)+1\approx2\kB T/\hbar\omega+1$, where $n_T(\omega)$ is the Bose-Einstein distribution function. The interaction with a structure that is infinitely extended in a transverse direction with respect to the e-beam (right) produces a divergence as $\Gamma(\omega)\propto1/\omega$ at $T=0$ and $\propto1/\omega^2$ at finite $T$.}
\label{Table1}
\end{table}

Because the loss of coherence relates to the different excitation amplitudes associated with each of the electron paths, it is pertinent to recall that the probability that a moving electron undergoes inelastic energy exchanges when passing near an extended material structure presents an infrared divergence due to the contribution of radiative modes \cite{paper127}, although we anticipate that such a divergence does not lead to any relevant physical pathology as the total energy loss and the degree of coherence remain both finite (see Appendix~\ref{divergencesinGamma}). The results summarized in Table~\ref{Table1} show that divergences are found when free electrons couple to extended material structures, for which the loss probability scales as $\Gamma(\omega)\propto1/\omega$ and $\propto1/\omega^2$ with the energy loss $\hbar\omega$ at zero and finite temperatures, respectively. We show in this work that these divergences produce a radical depletion of coherence for electron paths separated by large distances (i.e., when one of the paths is more exposed to the noted divergences), but a finite degree of coherence is always preserved, and full coherence is recovered as the path separation is reduced. Such preservation of coherence is a key element in off-axis electron holography \cite{LL02,WBT18}, which relies on interference between electrons passing either through or outside a material to reconstruct its atomic structure.

A two-path electron in which one of the paths is close to an extended structure should be a good example to observe a large degree of electron decoherence produced by radiative coupling. We thus consider the configurations depicted in Fig.~\ref{Fig1}a,b, and indeed, based on the rigorous theory presented below, we obtain a substantial increase in decoherence when one of the paths is placed 100~$\mu$m away from the edge of a perfect-electric-conductor (PEC) half-plane and the other path is separated by a distance of either 50~$\mu$m or 5~mm. This effect can be visualized through the interference fringes formed when the two paths are recombined (see Fig.~S1 in SI), as we show in Fig.~\ref{Fig1}c. A similar effect is observed while maintaining a large inter-path distance (a few mm) by placing a half-plane close (e.g., 1~$\mu$m) or far (10~$\mu$m) from the nearest electron path (see Supplementary Fig.~\ref{FigS2}). These are situations in which quantum-mechanical effects (decoherence) take place over large distances, a territory that was so far reserved to the lossless propagation of photons in free space \cite{HHH09,TCT10} or superpositions of matter states at low temperatures \cite{MMK96,BHD96}.

\begin{figure*}[th!]
\centering{\includegraphics[width=1.0\textwidth]{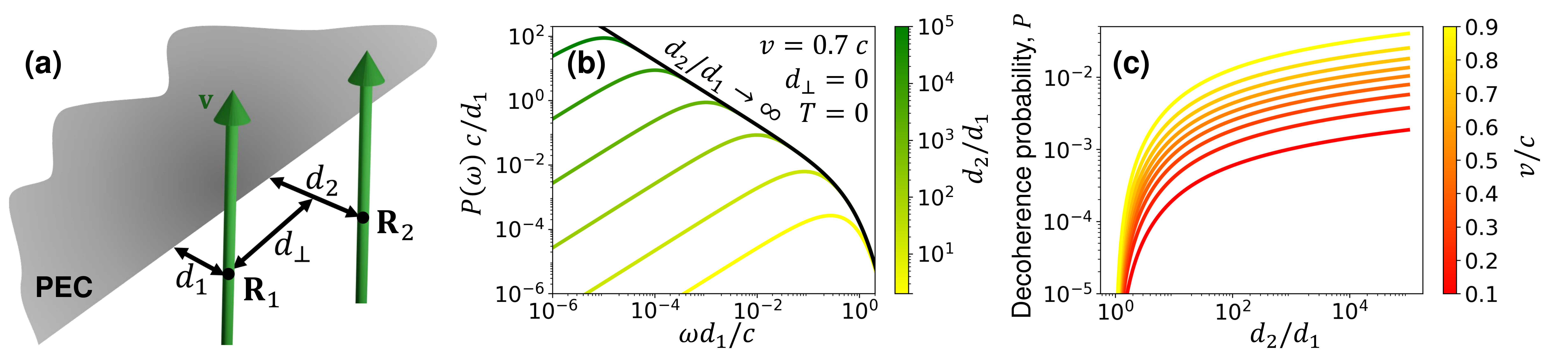}}
\caption{{\bf Two-path decoherence by a half-plane at zero temperature.} (a)~System under consideration, consisting of a single electron split into a two-path spatial superposition and passing close and perpendicularly to a perfectly conducting half-plane at distances $d_1$ and $d_2$ from the edge [beam path positions $\Rb_1=(0,d_1)$ and $\Rb_2=(d_\perp,d_2)$]. (b)~Universal plot of the spectrally resolved decoherence probability for various $d_2/d_1$ ratios (color curves) with $d_\perp=0$, approaching a divergent profile in the $d_2\gg d_1$ limit (black curve). We consider an electron velocity $v=0.7\,c$ and normalize the frequency and the probability using the smallest distance $d_1$. (c)~Decoherence probability as a function of $d_2/d_1$ for $d_\perp=0$ and different electron velocities (see color scale).}
\label{Fig2}
\end{figure*}

Here, we theoretically demonstrate that the presence of an extended material structure can produce strong electron decoherence on electron beams (e-beams) placed at an arbitrarily large distance from the material. We consider radiative modes of commensurably large wavelengths, for which the materials behave either as real-permittivity dielectrics or lossless perfect electric conductors (PECs), such that material excitations can be ignored. Specifically, we consider a PEC thin half-plane and a two-path e-beam passing perpendicularly to it (Fig.~\ref{Fig2}a). Because the half-plane in the zero-thickness limit is a scale-invariant structure and the PEC response eliminates any absolute length scale from the problem at zero temperature, we find that the decoherence between the two paths only depends on the ratio of their distances to the half-plane, and consequently, decoherence is predicted to take place for arbitrarily large macroscopic electron--half-plane distances, provided the inter-path separation is sufficiently large. At finite vacuum temperature $T$, the thermal wavelength $\lambda_T=2\pi\hbar c/\kB T$ plays a role by imposing an absolute length scale that is inversely proportional to $T$ (e.g., $\lambda_T\sim14$~mm at 1~K and 50~$\mu$m at room temperature). We find that decoherence is then boosted for large inter-path separations compared with $\lambda_T$, provided one of the paths passes near the half-plane. In a more practical scenario, we consider a finite-width ribbon and show that the half-plane limit is recovered for large width compared with the electron--edge distance. Our results support the use of electron decoherence to sense the presence of distant objects and measure the vacuum temperature.

\section{Results and discussion}

\subsection{General theory of electron-beam decoherence}

We are interested in investigating the loss of coherence among different spatial regions of a single electron prepared in a beam moving along $z$ with velocity $v$. The electron state can change due to the interaction with the environment (i.e., any material structure and the radiation field), giving rise to inelastic components that are position-dependent and, thus, decreasing the degree of coherence between separate spatial regions of the beam. As a practical manifestation of this effect, after propagation from those regions to an electron detector, the loss of coherence produces a reduction in the visibility of the resulting interference fringes, which we investigate here in a rigorous quantitative manner.

Describing an incident electron through its interaction-picture density matrix $\rho_e^{i}(\rb,\rb')$, scattering by a structure produces a final density matrix given by
\begin{align}
\rho_e^{f}(\rb,\rb') = \ee^{-P(\rb,\rb')+\ii \chi(\rb,\rb')}\rho_e^{i}(\rb,\rb'),
\label{evolutionFinal}
\end{align}
where
\begin{align}
P(\Rb,\Rb') &=\frac{1}{2}\int_0^{\infty} d\omega \big[2n_T(\omega)+1\big] \label{decoP}\\
&\times\bigg[ \Gamma(\Rb,\Rb,\omega) + \Gamma(\Rb',\Rb',\omega) - 2\Gamma(\Rb,\Rb',\omega) \bigg]
\nonumber
\end{align}
is the decoherence probability, which is in turn expressed as a frequency integral of the generalized loss probability
\begin{align}
\Gamma(\Rb,\Rb',\omega) = \frac{4e^2}{\hbar} \int_{-\infty}^{\infty} dz &\int_{-\infty}^{\infty} dz' \cos\Big[\frac{\omega}{v}(z-z')\Big] \label{probEindbis}\\
&\times\Imm\{-G_{zz}(\rb,\rb',\omega)\}.
\nonumber
\end{align}
(A self-contained derivation of these expressions is presented in Appendices~\ref{general} and \ref{secboson}). Here, the temperature $T$ enters through the Bose-Einstein distribution function $n_T(\omega)=\big(\ee^{\hbar\omega/\kB T}-1\big)^{-1}$, while the scattering structure is accounted for through the electromagnetic Green tensor $G(\rb,\rb',\omega)$, which can be calculated by solving the macroscopic electromagnetic response according to $\nabla \times \nabla \times G(\rb,\rb',\omega)-(\omega^2/c^2)\epsilon(\rb,\omega) G(\rb,\rb',\omega)=(-1/c^2)\delta(\rb -\rb')$ for any structure defined by a local, frequency-dependent permittivity $\epsilon(\rb,\omega)$. The real phase $\chi(\rb,\rb')$ in Eq.~(\ref{evolutionFinal}) is also expressed in terms of the Green function (see Appendix~\ref{secboson}), adding a rigid shift to the fringes observed in two-path interference. Following pioneering studies of decoherence in free-space electrons \cite{F93}, explicit results analogous to these expressions have been obtained by using macroscopic quantum electrodynamics \cite{SB12,paper357}, but the derivation that we present in the SI is self-contained and formulated in more general terms. Incidentally, the decoherence probability $P$ can take values larger than 1 since it must be understood as a depletion of coherence given by $\ee^{-P}$ according to Eq.~(\ref{evolutionFinal}). Analogously, the classical EELS probability can also exceed unity and must be understood as the mean of a Poissonian distribution of multiple losses \cite{paper228}.

\subsection{Electron decoherence by a half-plane}

The application of Eqs.~(\ref{decoP}) and (\ref{probEindbis}) to an e-beam passing outside and perpendicularly to a PEC half-plane produces analytical expressions for the decoherence probability, as shown in the self-contained derivation offered in the Appendix~\ref{appendixhalfplane}. More precisely, referring to the geometry depicted in Fig.~\ref{Fig2}a, involving two paths with transverse coordinates $\Rb_1=(d_1,0)$ and $\Rb_2=(d_2,d_\perp)$, we find
\begin{widetext}
\begin{align}
P(\Rb_1,\Rb_2)=&\frac{\alpha}{2\pi}
\int_0^1 \frac{d\mu}{\sqrt{1-\mu^2}}\;\frac{\big[(1+v^2/c^2)\mu^2+\eta^2\big]}{\big(\mu^2+\eta^2\big)^{3/2}}
\int_0^\infty \frac{d\theta}{\theta} \;\coth(\theta/4\pi) \label{PRRhalfplane}\\
&\times \bigg[ \ee^{-2\theta\,(d_1/\lambda_T)\sqrt{\mu^2+\eta^2}} + \ee^{-2\theta\,(d_2/\lambda_T)\sqrt{\mu^2+\eta^2}}
-2\cos\big(\mu\theta\,d_\perp/\lambda_T\big)\,\ee^{-\theta\,[(d_1+d_2)/\lambda_T]\sqrt{\mu^2+\eta^2}}\bigg], \nonumber 
\end{align}
\end{widetext}
where $\alpha = e^2/\hbar c\approx1/137$ is the fine structure constant and $\eta=c/v\gamma$ is a velocity-dependent parameter that uses the relativistic Lorentz factor $\gamma=1/\sqrt{1-v^2/c^2}$. The integration variable $\theta=2\pi\hbar\omega/\kB T$ in Eq.~(\ref{PRRhalfplane}) encapsulates the exchanged energy $\hbar\omega$, and we have rewritten the thermal factor as
\begin{align}
2n_T(\omega)+1=\coth(\theta/4\pi).
\label{nnp1} 
\end{align}
As we argue above, this factor and $\Gamma(\Rb,\Rb',\omega)$ are both diverging as $1/\omega$ in the $\omega\to0$ limit. However, the divergence is canceled because the expression inside curly brackets in Eq.~(\ref{PRRhalfplane}) behaves as $\propto\theta^2\sim\omega^2$ for small $\omega$: the first two terms inside the curly brackets represent the contributions arising from the two separate electron paths passing by $\Rb_1$ and $\Rb_2$, respectively, whereas the rightmost term stands for path interference, and while the integral of each of these three terms diverges, their sum becomes finite. Consequently, the decoherence probability $P(\Rb_1,\Rb_2)$ remains finite. We note that this quantity vanishes for $\Rb_1=\Rb_2$, as expected from Eq.~(\ref{decoP}), and it depends on $\Rb_1$, $\Rb_2$, and $T$ only through the ratios $d_1/\lambda_T$, $d_2/\lambda_T$, and  $d_\perp/\lambda_T$.

\subsubsection{Zero-temperature limit}

In the zero-temperature limit, we have $n_T(\omega)\to0$, so we can approximate $\coth(\theta/4\pi)\approx1$ in Eq.~(\ref{PRRhalfplane}) [see Eq.~(\ref{nnp1})]. The $\theta$ integral can then be performed analytically by first absorbing the $(1/\lambda_T)\sqrt{\mu^2+\eta^2}$ factor of the exponentials into the integration variable, and then considering the identity $\int_0^\infty d\theta\,\ee^{-a \sqrt{\theta^2+g^2}}/\sqrt{\theta^2+g^2}=K_0(ga)$ (see Eq.~3.914-4 in Ref.~\cite{GR1980}) together with the expansion $K_0(ga) = \log(2)-\log(ga)-\mathcal{C} + \mathcal{O}[(ga)^2\log(ga)]$ for $ga\ll1$, where $\mathcal{C}$ is the Euler constant. Applying this result to the three terms inside the curly brackets of Eq.~(\ref{PRRhalfplane}), setting $d_\perp=0$, and taking the $g\to0$ limit, we find
\begin{align}
P_{T=0}(d_1,d_2,d_\perp=0)=\frac{\alpha}{2\pi}\,f(v/c)\,\log\!\bigg[\frac{(d_1+d_2)^2}{4d_1d_2}\bigg],
\label{PT0simple}
\end{align}
where $f(v/c)=\int_0^1 d\mu\;\big[(1+v^2/c^2)\mu^2+\eta^2\big]\,(1-\mu^2)^{-1/2}$ $\times\big[\mu^2+\eta^2\big]^{-3/2}$ encapsulates the dependence on electron velocity via the variable $\eta=c/v\gamma$. Incidentally, this function admits the closed-form expression $f(\beta)=\beta\gamma\,(1+\beta^2)\, \mathcal{K}(-\beta^2\gamma^2)-(\beta^3/\gamma)\, \mathcal{E}(-\beta^2\gamma^2)$ in terms of the elliptical integrals $\mathcal{K}$ and $\mathcal{E}$. Interestingly, the dependences on path positions and electron velocity are factorized in Eq.~(\ref{PT0simple}). The distance dependence of the decoherence probability exhibits a logarithmic divergence as $P(d_1,d_2)\approx(\alpha/2\pi)\,f(v/c)\,|\!\log(d_2/d_1)|$ in the $d_2/d_1\to0,\infty$ limits, while it vanishes for $d_1=d_2$. In addition, $P(d_1,d_2)$ vanishes at $v=0$ and diverges as $\propto\log\gamma$ as the electron velocity approaches the speed of light.

It is instructive to examine the frequency integral of Eq.~(\ref{PRRhalfplane}) in the $T=0$ limit. For $d_\perp=0$ and $d_2>d_1$, Fig.~\ref{Fig2}b shows that low frequencies become increasingly relevant as we increase $d_2$, eventually converging to a profile that diverges as $\propto1/\omega$ at low frequencies in the $d_2/d_1\rightarrow\infty$ limit, for which the frequency integral is consequently infinite (i.e., we have full decoherence preventing any interference when mixing the two paths). We remark that an arbitrarily large loss of coherence can take place even when $d_1$ is made arbitrarily large, provided $d_2/d_1\gg1$, as the electron can always couple to long-wavelength excitations.

Universal curves for the decoherence probability are obtained from Eq.~(\ref{PT0simple}) for $d_\perp=0$ as a function of $d_2/d_1$ (Fig.~\ref{Fig2}c) for different electron velocities. Despite the logarithmic divergence with $d_2/d_1$ and the $\propto\log\gamma$ divergence as $v$ approaches $c$, the decoherence probability takes relatively small values at $T=0$ within the wide range of distances and velocities explored in Fig.~\ref{Fig2}c. This conclusion is however dramatically changed at finite temperatures, as we show below.

Similar results as those presented in Fig.~\ref{Fig2}b,c are obtained for the zero-temperature decoherence probability when varying the inter-path distance $d_\perp$ along the direction parallel to the half-plane edge while setting $d_1=d_2$ (see Supplementary Fig.~\ref{FigS3}), which we calculate by numerically integrating Eq.~(\ref{PRRhalfplane}) after setting $\coth(\theta/4\pi)=1$.

\begin{figure*}[th!]
\centering{\includegraphics[width=1.0\textwidth]{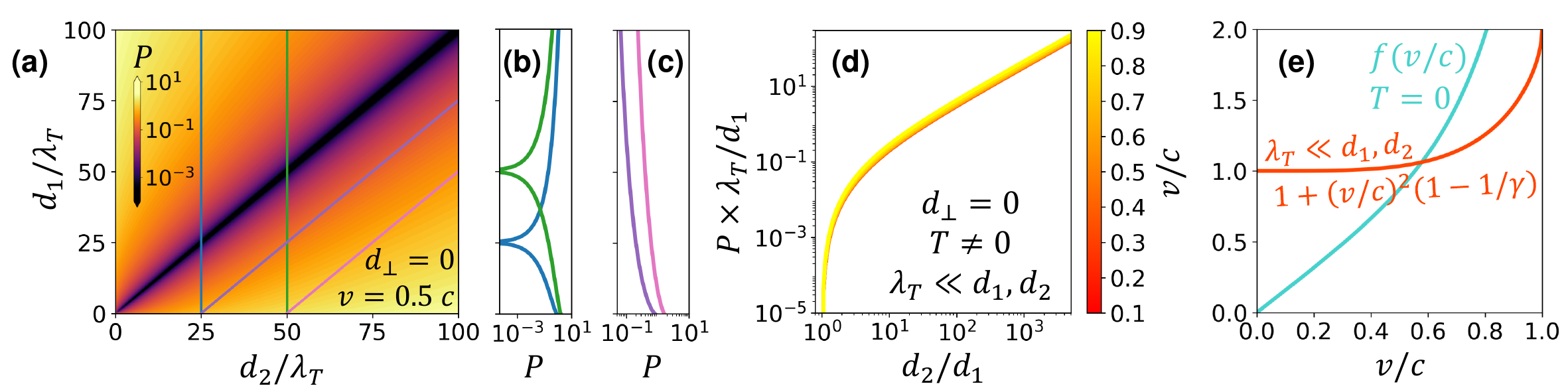}}
\caption{{\bf Decoherence at finite temperature.} (a) Decoherence probability under the two-path configuration of Fig.~\ref{Fig2}a as a function of path-edge distances $d_1$ and $d_2$, normalized to the thermal wavelength $\lambda_T=2\pi c\hbar/\kB T$. We set the electron velocity to $v=0.5\,c$. (b,c) Cuts of the probability in (a) along the color-coordinated lines. (d) Decoherence probability in the high-temperature limit ($d_1, d_2\gg\lambda_T$), in which $\lambda_T P/d_1$ is only a function of $d_2/d_1$. (e) Electron-velocity prefactors in the $T=0$ and $d_1, d_2\gg\lambda_T$ limits. We set $d_\perp=0$ in all cases.}
\label{Fig3}
\end{figure*}

\subsubsection{Decoherence at finite temperature}

We examine the full dependence of the decoherence probability $P$ [Eq.~(\ref{PRRhalfplane})] on $d_1/\lambda_T$ and $d_2/\lambda_T$ for $d_\perp=0$ in Fig.~\ref{Fig3}a-c, setting $v/c=0.5$ as an illustrative example since the dependence on velocity is relatively mild (see Supplementary Fig.~\ref{FigS4}). The diagonal of the plot in Fig.~\ref{Fig3}a is dominated by a substantial reduction in the decoherence probability when $|d_1-d_2|\lesssim\lambda_T$ (see also Fig.~\ref{Fig3}b; also $P=0$ for $d_1=d_2$). However, $P$ quickly rises to large values when the distance difference is a few times the thermal wavelength (Fig.~\ref{Fig3}c).

It is interesting to analytically examine the high-temperature limit, in which the integral in Eq.~(\ref{PRRhalfplane}) is dominated by regions where $n_T(\omega)=\approx\kB T/\hbar\omega\gg1$, so we can approximate $\coth(\theta/4\pi)\approx4\pi/\theta$ [see Eq.~(\ref{nnp1})]. Setting again $d_\perp=0$, and changing the $\theta$ variable of integration to absorb the $(1/\lambda_T)\sqrt{\mu^2+\eta^2}$ factor, the $\theta$ integral can be analytically performed by using the identity  $\int_0^\infty d\theta\,\ee^{-a\theta}/(\theta^2+g^2)=(1/g)[{\rm Ci}(ga)\sin(ga)-{\rm Si}(ga) \cos(ga)]$ (see Eq.~3.354-1 in Ref.~\cite{GR1980}), where ${\rm Ci}$ and ${\rm Si}$ are the cosine and sine integral functions, respectively. We then expand ${\rm Ci}(ga) = \mathcal{C} + \log(ga)+\mathcal{O}(ga)^2$ and ${\rm Si}(ga) = -\pi/2+ga+\mathcal{O}(ga)^2$ for $ga\ll1$, and from here, we find $\int_0^\infty d\theta\,\ee^{-a\theta}/(\theta^2+g^2)\approx a\log a+\cdots$, where the eliminated terms are linear in $a$, independent of $a$, or vanishing in the $g\to0$ limit, so they do not contribute to the $\theta$ integral when summing the three exponential terms in Eq.~(\ref{PRRhalfplane}). The $\mu$ integral of the remaining $a\log a$ contribution can also be performed in closed form, leading to the final result
\begin{align}
&P_{d_1,d_2\gg\lambda_T}(d_1,d_2,d_\perp=0) = 2\pi\alpha \big[1+(v^2/c^2)(1-1/\gamma)\big] \nonumber\\
&\quad\quad\times\bigg[\frac{d_1}{\lambda_T}\log\Big(\frac{2d_1}{d_1+d_2}\Big) + \frac{d_2}{\lambda_T}\log\Big(\frac{2d_2}{d_1+d_2}\Big)\bigg], \label{PoT}
\end{align}
where we again observe a factorization of the dependences on electron--half-plane distances and electron velocity. The decoherence probability $P$ exhibits a linear divergence with $d_1$ and $d_2$ for a constant ratio $d_2/d_1$ in the $d_1,d_2\gg\lambda_T$ limit. In addition, the temperature enters through an overall factor $1/\lambda_T=\kB T/2\pi\hbar c\propto T$, so that $P$ also scales linearly with $T$.

In the high-temperature limit [Eq.~(\ref{PoT})], the scaled probability $\lambda_T P/d_1$ only depends on the ratios $d_2/d_1$ and $v/c$, and in particular, it exhibits a roughly linear increase with $d_2/d_1$, as shown in Fig.~\ref{Fig3}d. We further observe the noted linear scaling with $T$, directly reflecting the linear increase with temperature in the photon population at long photon wavelengths (i.e., those that are commensurate with the electron--edge distances, which are large compared with $\lambda_T$ in the limit under examination). In addition, the dependence on electron velocity is fully contained in the prefactor $1\le1+(v^2/c^2)(1-1/\gamma)\le2$ in Eq.~(\ref{PoT}), which takes finite values over a broad range of velocities typically used in electron microscopes, down to $v=0$ (Fig.~\ref{Fig3}e). This is in contrast to the $T=0$ behavior, in which, although $P$ also depends on velocity through a prefactor $f(v/c)$ [see Eq.~(\ref{PT0simple})], the latter vanishes in the small velocity limit.

We stress that the change in behavior from zero to finite temperature is continuous but relatively steep, as shown in Supplementary Fig.~\ref{FigS5}.

\begin{figure*}[th!]
\centering{\includegraphics[width=0.60\textwidth]{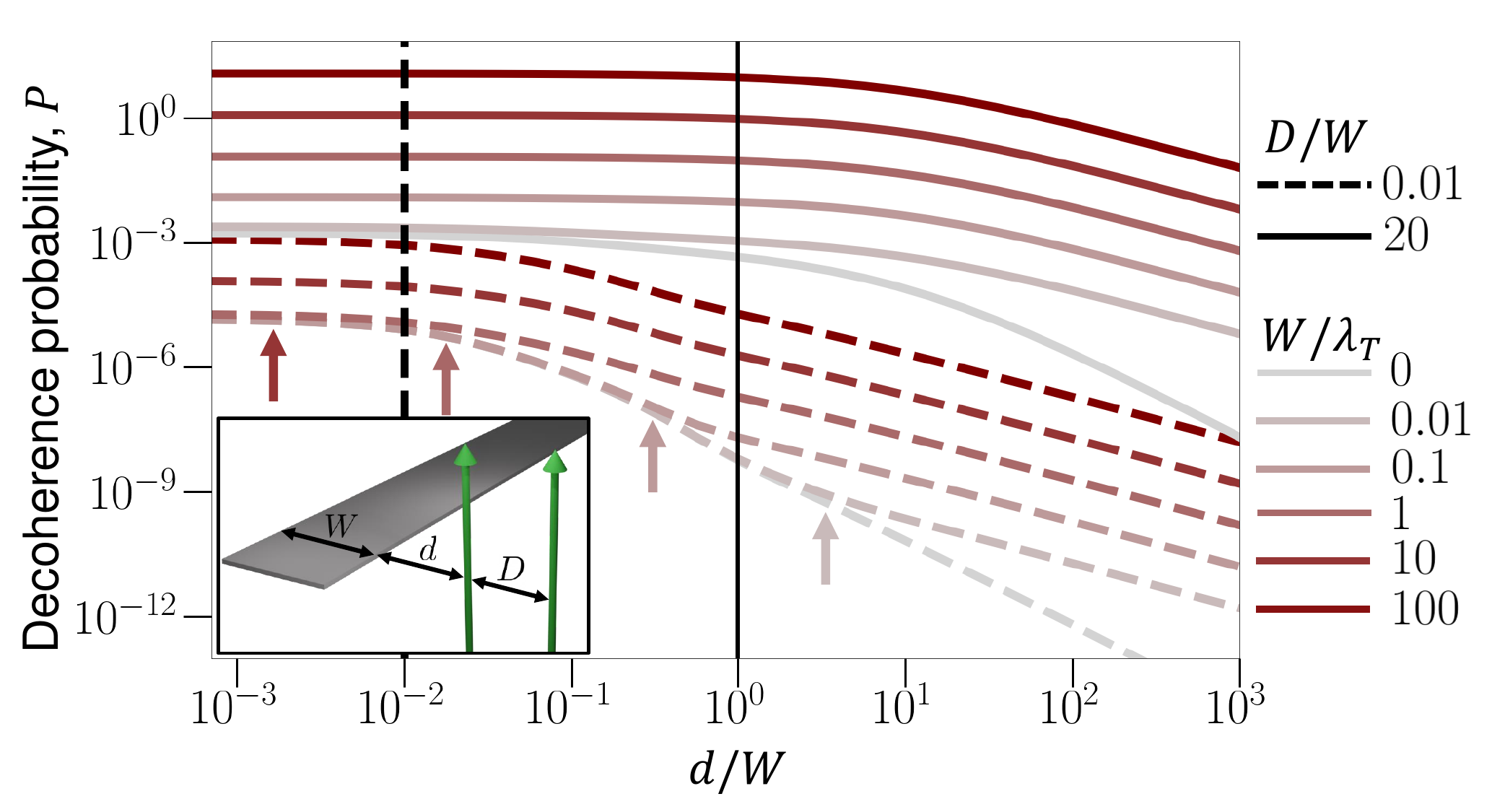}}
\caption{{\bf Finite size effects on the decoherence probability.} We plot the decoherence probability $P$ for the system sketched in the lower-left inset. The e-beam is prepared in a two-path superposition state (inter-path distance $D$) and interacts with a thin PEC ribbon of width $W$. The path-ribbon distances are $d$ and $d+D$. We present results for different values of the $D/W$ and $W/\lambda_T$ ratios (see the legend on the right).}
\label{Fig4}
\end{figure*}

\subsection{Finite-size effects: decoherence by a metallic ribbon}

While the assumption of a thin PEC screen is reasonable for metallic films of small thickness compared with the electron--edge distances, a finite extension of the half-plane geometry can play a role because the aforementioned infrared divergence requires that the structure responds at arbitrarily low frequencies (see Table~\ref{Table1}). We study finite-size effects by limiting the extension of the half-plane in one direction and considering instead a ribbon of finite width $W$. The decoherence probability is then computed by employing an ad hoc boundary-element method in which the ribbon is discretized through a uniform set of points along the transverse direction, as explained in the self-contained Appendix~\ref{ribbon}.

The resulting decoherence probability is plotted in Fig.~\ref{Fig4} for a two-path configuration featuring an inter-path distance $D$ and a shortest electron--ribbon distance $d$ (see inset and Supplementary Fig.~\ref{FigS6}). Specifically, we show calculations for $D/W=0.01$ and 20, combined with different $W/\lambda_T$ ratios ranging from zero temperature to $W=\lambda_T$. At large inter-path separations ($D=20\,W$, solid curves), the infinite half-plane limit is recovered for distances $d\ll W$. The condition $d=W$ (vertical solid line) signals the transition between the half-plane limit and a regime in which the probability is exponentially attenuated when increasing $d$ at all temperatures. This behavior is produced by ribbon-mediated coupling of the path that is closest to the edge to radiative modes, while the distant path experiences a negligible degree of inelastic interaction.

For relatively small inter-path separations ($D=0.01\,W$, dashed curves in Fig.~\ref{Fig4}), both paths undergo a similar level of inelastic interaction, and consequently, $P$ is strongly reduced compared to the results for large $D$. Under these conditions, the half-plane limit is recovered at $d\sim D=0.01\,W$ (vertical dashed line) in both the low- and high-temperature regimes ($W/\lambda_T=0$ and $\gg1$, respectively). Similarly to the half-plane, a departure from the $T=0$ regime is observed at large ribbon--electron separations, as indicated by color-coordinated vertical arrows in Fig.~\ref{Fig4}. At large temperatures, this departure takes place over the entire range of distances $d$ considered in the figure, thus producing an overall increase in the decoherence probability.

\section{CONCLUSIONS}

In summary, inelastic radiative scattering of free electrons passing near extended structures produces a divergence in electron decoherence at high temperatures and/or large inter-path separations for electrons prepared in a two-path beam configuration. In essence, the extended material structure acts as a coupler between the evanescent electron field and long-wavelength free-space radiation. We exemplify this effect through an analytical treatment of the interaction between free electrons and a metallic half-plane, which is a self-scaling geometry such that, at zero temperature, there are no absolute length scales in the system, and therefore, the decoherence probability only depends on the ratio of electron-path distances to the edge. For this system, the probability that the electron interacts with radiative modes receives a divergent contribution at low frequencies and, although this yields an infinite frequency-integrated probability, the loss of coherence remains finite for finite electron inter-path separation. The decoherence probability increases with temperature as we depart from $T=0$. Then, the thermal wavelength defines an absolute length scale in the system. These results require the involvement of low-frequency radiation, with wavelengths that are commensurate with both the electron-path--edge distances and the extension of the material, as confirmed by the observation that the half-plane limit is recovered when considering instead ribbons of large width compared with the thermal wavelength and the electron--edge distance.

These results suggest the possibility of detecting the presence of distant objects without perturbing them (i.e., without causing any inelastic excitation in the involved materials, and relying instead on decoherence produced by coupling to radiative modes). Indeed, at zero temperature, the self-scaling nature of the half-plane geometry implies that a large decoherence probability is obtained for any arbitrarily large electron--edge distance, provided the latter is small compared with the inter-path separation. In addition, at finite temperatures, a high degree of decoherence is observed when the inter-path distance is large compared with the thermal wavelength.

The strong temperature dependence of the decoherence probability could be exploited to perform vacuum thermometry and measure the temperature of the free-space thermal radiation bath. The required inter-path distances are a few times the thermal wavelength. At room temperature, the latter is $\lambda_T\approx50$~$\mu$m, so we need to consider distances of hundreds of microns, which are typical separations between e-beams and different structural components in electron microscopes. Incidentally, some degree of undesired electron decoherence could be produced due to radiative coupling assisted by elements placed close to the e-beam in electron microscopes, an effect that deserves further examination in light of the results presented in this work.

Our predictions could therefore be tested in an electron microscope by introducing a specimen consisting of a wide ribbon (e.g., having width $W=20\,\lambda_T\approx1$~mm at room temperature) and splitting the e-beam into two paths (e.g., separated by a distance $D=W$). Then, interference between the two paths is significantly reduced when bringing the ribbon within a distance $d<0.1\,W\approx0.1$~mm from the nearest electron path (e.g., $P>0.2$). We further conceive a macroscopic version of this experiment at cryogenic temperatures, for which the decoherence probability is preserved if all lengths are scaled by the thermal wavelength (e.g., multiplied by a factor of $\sim100$ when moving from room temperature to outer-space thermal-background conditions at 2.7~K). The far-field interference arising when mixing electron paths that are separated by a distance $D$ results in interference fringes with an angular spacing $\propto1/D$, which becomes too small to be experimentally resolved at macroscopic separations $D$ of hundreds of microns. Instead, an electron optics system could be used to split an e-beam and separate the electron paths to the desired distance $D$ in the region of interaction with the material structure, followed by a second set of optical components that bring the electron paths to interference at the post-selecting transmission grating \cite{JTM21,paper388} (see Supplementary Fig.~\ref{FigS1}).

The $1/r$ distance dependence of the electromagnetic fields that accompany a moving charge (the electron) underscores the observed divergences in electron decoherence. An analogous divergence in decoherence could potentially be produced by other types of excitations that share similar long-range behavior. In particular, further investigation is needed to explore the effect of coupling between massive particle waves and gravitons, as well as the gravitational interaction with long-range modes in material structures (e.g., sound and elastic waves).


\acknowledgments
We thank Archie Howie and Morgan Mitchell for helpful and enjoyable discussions. This work has been supported in part by the European Research Council (Advanced Grant 789104-eNANO), the European Commission (Horizon 2020 Grants No. 101017720 FET-Proactive EBEAM and No. 964591-SMART-electron), the Spanish MICINN (PID2020-112625GB-I00 and Severo Ochoa CEX2019-000910-S), the Catalan CERCA Program, and Fundaci\'{o}s Cellex and Mir-Puig.

\appendix
\begin{widetext}
\section*{APPENDICES}

We present a self-contained derivation of the theory of electron decoherence based on first principles, along with supplementary results for a two-path electron beam oriented parallel to a half-plane edge, the spectral decomposition of the decoherence probability, the path-separation dependence of the decoherence probability produced by interaction with a ribbon, and a proposal for measuring electron interference fringes compatible with having a large inter-path separation at the position of the specimen.

\renewcommand{\thesection}{A} 
\renewcommand{\theequation}{A\arabic{equation}} 
\section{General description of electron-beam decoherence}
\label{general}

Starting from the Dirac equation to describe the electron and the radiation field, working in an electromagnetic gauge with zero scalar potential (the so-called temporal gauge), neglecting ponderomotive interactions, and adopting the nonrecoil approximation (i.e., considering an electron of high kinetic energy $E_0$ and constant electron velocity $\vb$ under the assumption of small momentum exchanges with the environment compared with the initial electron momentum $\hbar\qb_0$), one finds the approximate real-space Hamiltonian $\HH(\rb)=\HH_{\rm rad}+\HH_{\rm el}(\rb)+\HH_{\rm int}(\rb)$, where $\HH_{\rm rad}$ and $\HH_{\rm el}(\rb)=E_0-\hbar\vb\cdot\qb_0-\ii\hbar\vb\cdot\nabla$ describe the noninteracting radiation and electron components, while $\HH_{\rm int}(\rb)=(e\vb/c)\cdot\Abh(\rb)$ is the minimal-coupling radiation--electron interaction, which is simply proportional to the vector potential operator $\Abh(\rb)$ (see Ref.~\cite{paper339} for a detailed derivation starting from the Dirac equation). It is useful to move to the interaction picture, in which the interaction Hamiltonian becomes
\begin{align}
\Hint(\rb,t)&=\ee^{\ii[\HH_{\rm rad}+\HH_{\rm el}(\rb)]t/\hbar}\,\HH_{\rm int}(\rb)\,\ee^{-\ii[\HH_{\rm rad}+\HH_{\rm el}(\rb)]t/\hbar} \nonumber\\
&=\frac{e\vb}{c}\cdot \Abh(\rb+\vb t,t),
\label{Hint1}
\end{align}
where $\Abh(\rb)$ inside $\HH_{\rm int}(\rb)$ acquires a time dependence from both the transformation $\Abh(\rb,t)=\ee^{\ii\HH_{\rm rad}t/\hbar}\Abh(\rb)\ee^{-\ii\HH_{\rm rad}t/\hbar}$ and the displacement introduced by the $-\ii\hbar\vb\cdot\nabla$ term in $\HH_{\rm el}$ [i.e., we apply the identity \cite{notedecoherence1}
%
$\ee^{\vb t\cdot\nabla}f(\rb)\ee^{-\vb t\cdot\nabla}g(\rb)=f(\rb+\vb t)g(\rb)$, which is valid for any $\rb$-dependent functions $f(\rb)$ and $g(\rb)$].

It is convenient to describe the temporal dynamics through the evolution operator $\hat{\mathcal{S}}(\rb,t)$, defined in such a way that the combined electron--environment state at time $t$ is given by $\ket{\psi(\rb,t)}=\hat{\mathcal{S}}(\rb,t)\ket{\psi(\rb,-\infty)}$, where $\ket{\psi(\rb,-\infty)}$ is the state in the infinite past and a ket notation is adopted to indicate the radiation degrees of freedom. To study the evolution of the electron, we propagate the full density matrix using $\hat{\mathcal{S}}(\rb,t)$ and eventually trace out the radiation. We consider the initial state of the electron--environment system to be uncorrelated [i.e., expressed as a tensor product of an electron density matrix $\rho^i_{e}(\rb,\rb')$ and a thermal distribution of the environment degrees of freedom]. In addition, we neglect any memory effect in the environment, such that it can be approximated as a thermal state of constant temperature $T$ at all times. Putting these elements together, the time-dependent electron-projected density matrix reads
\begin{align}
\dens(\rb,\rb',t)=\big\langle\hat{\mathcal{S}}(\rb,t)\hat{\mathcal{S}}^\dagger(\rb',t)\big\rangle_T\,\rho_e^{i}(\rb,\rb'),
\label{dssd1}
\end{align}
where $\langle\cdots\rangle_T$ indicates the thermal average over the radiation field. At any time after the interaction has taken place, this expression reduces to
\begin{align}
\rho_e^{f}(\rb,\rb') = \ee^{-P(\rb,\rb')+\ii \chi(\rb,\rb')}\rho_e^{i}(\rb,\rb')
\nonumber
\end{align}
[Eq.~(1) in the main text], where $P(\rb,\rb')$ and $\chi(\rb,\rb')$ are time-invariant quantitites that depend on $z$ and $z'$ only through the difference $z-z'$ for $\vb$ along $\zz$ (see below). Note that, in the interaction picture, the density matrix $\dens(\rb,\rb')$ does not evolve with time in the absence of any perturbation, while a trivial time dependence is acquired as $\dens(\rb-\vb t,\rb'-\vb t) = \ee^{-\ii\HH_{\rm el}(\rb)t/\hbar} \dens(\rb,\rb') \ee^{\ii\HH_{\rm el}(\rb')t/\hbar}$ in the Schr\"odinger picture, reflecting the translation due to the electron motion, which is computed again through the aforementioned displacement. This allows us to write
\begin{align}
\dens(\rb-\vb t,\rb'-\vb t) = \ee^{-P(\rb,\rb')+\ii \chi(\rb,\rb')}\rho_e^{i}(\rb-\vb t,\rb'-\vb t)
\label{rhoPchirho}
\end{align}
for the electron density matrix in the Schr\"odinger picture. Here, the real functions $P(\rb,\rb')$ and $\chi(\rb,\rb')$ are defined by
\begin{align}
\ee^{-P(\rb,\rb')+\ii \chi(\rb,\rb')}=\big\langle\hat{\mathcal{S}}(\rb,\infty)\hat{\mathcal{S}}^\dagger(\rb',\infty)\big\rangle_T,
\label{epchi1}
\end{align}
such that they account for spatially dependent decoherence and an elastic phase \cite{paper357}, respectively. To find more explicit expressions for $P(\rb,\rb')$ and $\chi(\rb,\rb')$, we use the Magnus expansion up to second order \cite{M1954} and approximate the evolution operator as
\begin{align} \label{Stt0}
\hat{\mathcal{S}}(\rb,t) \approx \exp\bigg\{&-\frac{\ii}{\hbar} \int_{-\infty}^{t} dt'\; \Hint(\rb,t') 
-\frac{1}{2\hbar^2}\int_{-\infty}^{t} dt'\int_{-\infty}^{t'} dt''\; \big[\Hint(\rb,t'),\Hint(\rb,t'')\big]\bigg\}.
\end{align}
Now, introducing Eq.~(\ref{Hint1}) into Eq.~(\ref{Stt0}), we can write $\hat{\mathcal{S}}(\rb,\infty)=\exp\{-\Bop(\rb) - (1/2) \Gop(\rb)\}$ in terms of the operators
\begin{subequations}
\label{BG}
\begin{align}
&\Bop(\rb) = \frac{\ii ev}{\hbar c}\int_{-\infty}^\infty dt\; \Ah_z(\rb-\vb t,t), \label{Bop}\\
&\Gop(\rb) = \Big(\frac{ev}{\hbar c}\Big)^2 \int_{-\infty}^\infty dt \int_{-\infty}^t dt'
\;\big[\Ah_z(\rb-\vb t,t), \Ah_z(\rb-\vb t',t')\big], \label{Gop}
\end{align}
\end{subequations}
where we set $\vb=v\,\zz$ without loss of generality. Then, using the Baker-Campbell-Hausdorff formula and retaining only terms up to second order in $\Abh$, Eq.~(\ref{epchi1}) becomes
\begin{align}
\ee^{-P(\rb,\rb')+\ii \chi(\rb,\rb')} \approx \bigg\langle\exp\Big(&-\Bop(\rb)+\Bop(\rb') -\frac{1}{2} \big\{[\Bop(\rb),\Bop(\rb')]+\Gop(\rb)-\Gop(\rb')\big\}\Big)\bigg\rangle_T,
\label{epchi2}
\end{align}
where we have employed the properties $\Bop^\dagger=-\Bop$ and $\Gop^\dagger=-\Gop$. Finally, we use the cumulant expansion $\big\langle\ee^{\lambda \hat{C}}\big\rangle_T = \exp\big(\sum_{m=1}^\infty \lambda^m C_m\big)$ to evaluate Eq.~(\ref{epchi2}) up to second order in $\Abh$, for which one needs to Taylor-expand both sides of this expression around $\lambda=0$, identify terms proportional to each power of the variable $\lambda$, and finally set $\lambda=1$. Following this procedure, we readily find $C_1=\langle \hat{C} \rangle_T$ and $C_2=(1/2)\big(\langle \hat{C}^2 \rangle_T-\langle\hat{C}\rangle_T^2\big)$. Setting $\hat{C}=-\Bop(\rb)+\Bop(\rb')-(1/2) \big\{[\Bop(\rb),\Bop(\rb')]+\Gop(\rb)-\Gop(\rb')\big\}$ in accordance with Eq.~(\ref{epchi2}), we finally obtain
\begin{align}
-P(\rb,\rb')+\ii \chi(\rb,\rb') \approx\;
&\frac{1}{2}\Big\langle
\big[ \Bop(\rb)-\Bop(\rb') \big]^2-\big[\Bop(\rb),\Bop(\rb')\big]-\Gop(\rb)+\Gop(\rb')\Big\rangle_T \nonumber\\
&-\big\langle\Bop(\rb)-\Bop(\rb')\big\rangle_T
-\frac{1}{2} \big\langle \Bop(\rb)-\Bop(\rb') \big\rangle_T^2
\label{epchi3}
\end{align}
to second order in the interaction. Equations~(\ref{BG}) and (\ref{epchi3}) can be applied to any vector potential incorporating the optical response of a material structure through the radiation modes in $\Abh(\rb,t)$. Importantly, Eqs.~(\ref{Stt0}), (\ref{epchi2}), and (\ref{epchi3}) become exact in a commonly encountered scenario when $\Abh(\rb,t)$ is constructed from bosonic operators (see Sec.~\ref{secboson}). Nevertheless, Eq.~(\ref{epchi3}) remains valid to second order even if the radiation modes involve coupling to non-bosonic excitations, such as those resulting from scattering by systems with a finite number of energy levels or when nonlinear response effects are significant.

\renewcommand{\thesection}{B} 
\renewcommand{\theequation}{B\arabic{equation}} 
\section{Decoherence by coupling to bosonic excitations}
\label{secboson}

In most practical situations, the radiation field is composed of bosonic modes representing free photons as well as polarization and scattering by material structures. Then, each mode $j$ contributes with a term $\alpha_j^{*}\aa_j^\dagger+\alpha_j\aa_j$ to the vector potential operator $\Abh_z(\rb,t)$, where $\aa_j^\dagger$ and $\aa_j$ are creation and annihilation operators, while $\alpha_j$ are spatiotemporally dependent complex coefficients. The bosonic commutation relations $[\aa_j,\aa^\dagger_{j'}]=\delta_{jj'}$ directly imply that $[\Hint(\rb,t),\Hint(\rb',t')]$ is a c-number, and consequently, we have $[\Hint(\rb,t),[\Hint(\rb',t'),\Hint(\rb'',t'')]]=0$. This result renders Eqs.~(\ref{Stt0}) and (\ref{epchi2}) exact, as all terms beyond second order (which we neglected in the Magnus expansion and the Baker-Campbell-Hausdorff formula) are proportional to commutators of three of more interaction terms.

Likewise, when computing the thermal average in Eq.~(\ref{epchi2}), we note that $\Gop(\rb)$ [Eq.~(\ref{Gop})] and $[\Bop(\rb),\Bop(\rb')]$ [see Eq.~(\ref{Bop})] are both commutators of vector potential operators and, therefore, c-numbers that can be pulled outside $\langle\dots\rangle_T$. In addition, $\Bop(\rb)-\Bop(\rb')$ becomes a sum over terms of the form $\beta_j^*\aa_j^\dagger-\beta_j\aa_j$, where $\beta_j$ are complex coefficients that depend on $\rb$ and $\rb'$. Using the identity \cite{notedecoherence2,notedecoherence3}
$\big\langle\ee^{c_j^*\aa_j^\dagger-c_j\aa_j}\big\rangle_T = \exp\big\{(1/2)\big\langle (c_j^*\aa_j^\dagger-c_j\aa_j)^2\big\rangle_T\big\}$ together with $\big\langle\big[\sum_j(c_j^*\aa_j^\dagger-c_j\aa_j)\big]^2\big\rangle_T=\big\langle\sum_j(c_j^*\aa_j^\dagger-c_j\aa_j)^2\big\rangle_T$ (i.e., cross terms vanish under the thermal average), we then obtain $\big\langle\ee^{-\Bop(\rb)+\Bop(\rb')}\big\rangle_T = \exp\big\{(1/2)\big\langle[\Bop(\rb)-\Bop(\rb')]^2\big\rangle_T\big\}$. From these considerations, we find that Eq.~(\ref{epchi2}) leads to the exact relation (for bosonic excitations)
\begin{align}
-P(\rb,\rb')+&\ii \chi(\rb,\rb') = \frac{1}{2}\bigg[\Big\langle\big[\Bop(\rb)-\Bop(\rb') \big]^2\Big\rangle_T -\big[\Bop(\rb),\Bop(\rb')\big] -\Gop(\rb)+\Gop(\rb')\bigg],
\label{epchi4}
\end{align}
with $\Bop(\rb)$ and $\Gop(\rb)$ defined in Eqs.~(\ref{BG}). This expression coincides with Eq.~(\ref{epchi3}) because $\langle\Bop(\rb)\rangle_T=0$ for these types of modes.

To evaluate Eq.~(\ref{epchi4}), we use the relation
\begin{align}
\langle \Ah_a(\rb,t) &\Ah_{a'}(\rb',t') \rangle_T = -4\hbar c^2 \int_0^\infty d\omega \; \Imm\big\{G_{aa'}(\rb,\rb',\omega)\big\} \; \bigg\{2n_T(\omega)\cos[\omega(t-t')]+\ee^{-\ii\omega(t-t')}\bigg\}
\label{AAaverages1}
\end{align}
for the thermal average of the product of two vector potential operators, which is self-consistently derived in Appendix~\ref{AppendixA}. Here, $a$ and $a'$ denote Cartesian components, we introduce the Bose-Einstein distribution function
\begin{align}
n_T(\omega) = \frac{1}{\ee^{\hbar\omega/\kB T}-1}
\label{nTw}
\end{align}
at temperature $T$, and $G(\rb,\rb',\omega)$ is the electromagnetic Green tensor, implicitly defined by the equation
\begin{align}
\nabla \times \nabla \times G(\rb,\rb',\omega)& -\frac{\omega^2}{c^2}\epsilon(\rb,\omega) \, G(\rb,\rb',\omega) =-\frac{1}{c^2}\delta(\rb -\rb'),
\label{Greentensor}
\end{align}
where $\epsilon(\rb,\omega)$ is the space- and frequency-dependent local permittivity. In practice, $G(\rb,\rb',\omega)$ can be calculated analytically in simple geometries (e.g., spherical and planar interfaces) or via numerical electromagnetic simulations. From Eq.~(\ref{AAaverages1}), using the reciprocity property $G_{aa'}(\rb,\rb',\omega)=G_{a'a}(\rb',\rb,\omega)$, we readily find
\begin{align}
&[\Ah_a(\rb,t),\Ah_{a'}(\rb',t')] = 8\ii\hbar c^2\int_0^\infty d\omega\; \sin\big[\omega(t-t')\big]\; \Imm\big\{G_{aa'}(\rb,\rb',\omega)\big\}, \label{AAaverages2}
\end{align}
where we have dismissed $\langle\cdots\rangle_T$ because the commutator is a c-number for bosonic operators (see above) and, therefore, independent of the state of the radiation field.

Using Eq.~(\ref{AAaverages2}), we can transform Eq.~(\ref{Gop}) into
\begin{align}
\Gop(\Rb)=&\frac{4e^2\,\ii}{\hbar} \int_{-\infty}^\infty dz\int_{-\infty}^\infty dz'\int_0^\infty d\omega
\;\cos\Big[\frac{\omega}{v}(z-z')\Big] \; \Ree\big\{G_{zz}(\Rb,z,\Rb,z',\omega)\big\}
\label{GGfinal}
\end{align}
(see Appendix~\ref{AppendixB}), where we have indicated that the result only depends on the transverse coordinates $\Rb=(x,y)$. Likewise, Eq.~(\ref{AAaverages1}) allows us to directly evaluate the $N_1$ terms in Eq.~(\ref{epchi4}) and finally write
\begin{subequations}
\label{chiPfinal}
\begin{align}
\chi(\rb,\rb')&=\frac{2e^2}{\hbar} \int_{-\infty}^\infty dz''\int_{-\infty}^\infty dz'''\int_0^\infty d\omega
\;\bigg\{2\sin\Big[\frac{\omega}{v}(z-z'-z''+z''')\Big]
\;\Imm\big\{G_{zz}(\Rb,z'',\Rb',z''',\omega)\big\} \nonumber\\
&\quad\quad\quad\quad\quad\quad+\cos\Big[\frac{\omega}{v}(z''-z''')\Big]
\;\Ree\big\{G_{zz}(\Rb',z'',\Rb',z''',\omega)-G_{zz}(\Rb,z'',\Rb,z''',\omega)\big\}
\bigg\}, \\
P(\rb,\rb')&= \frac{1}{2}\int_0^\infty d\omega
\;\big[2n_T(\omega)+1\big]
\;\big[\Gamma(\rb,\rb,\omega) +\Gamma(\rb',\rb',\omega) -2\,\Gamma(\rb,\rb',\omega) \big],
\label{Prrfinal}
\end{align}
\end{subequations}
where
\begin{align}
\Gamma(\rb,\rb',\omega) = \frac{4e^2}{\hbar} \int_{-\infty}^\infty dz''\int_{-\infty}^\infty dz'''
\;\cos\Big[\frac{\omega}{v}(z-z'-z''+z''')\Big]
\;\Imm\big\{-G_{zz}(\Rb,z'',\Rb',z''',\omega)\big\}. \label{GammaRR}
\end{align}
The decoherence probability in Eqs.~(\ref{Prrfinal}) and (\ref{GammaRR}) coincides with the result derived in Ref.~\cite{SB12}. For $z=z'$, these expressions reduce to
\begin{align}
&P(\Rb,\Rb') =\frac{1}{2}\int_0^{\infty} d\omega \big[2n_T(\omega)+1\big] \; \bigg[ \Gamma(\Rb,\Rb,\omega) + \Gamma(\Rb',\Rb',\omega) - 2\Gamma(\Rb,\Rb',\omega) \bigg]
\nonumber \\
&\Gamma(\Rb,\Rb',\omega) = \frac{4e^2}{\hbar} \int_{-\infty}^{\infty} dz \int_{-\infty}^{\infty} dz' \cos\Big[\frac{\omega}{v}(z-z')\Big] \; \Imm\big\{-G_{zz}(\rb,\rb',\omega)\big\}
\nonumber
\end{align}
[Eqs.~(2) and (3) in the main text]. The present analysis remains valid to any order of interaction for quasi-monochromatic electrons in general under the nonrecoil approximation, thus generalizing previous calculations for a classical electron \cite{paper357} and for a quantum electron within second-order perturbation theory \cite{paper371}. Reassuringly, these functions satisfy the properties $P(\rb,\rb)=\chi(\rb,\rb)=0$ (thus guaranteeing the conservation of the total electron probability), $P(\rb,\rb')=P(\rb',\rb)$, and $\chi(\rb,\rb')=-\chi(\rb',\rb)$ (i.e., reciprocity is inherited from the Green tensor). Incidentally, the latter (symmetry with respect to the exchange of $\rb$ and $\rb'$) could already be anticipated by applying the Hermiticity of the density matrix $\rho(\rb,\rb')=\rho^*(\rb',\rb)$ to Eq.~(\ref{rhoPchirho}). The present self-contained derivation relies just on the assumption of bosonic excitations and the nonrecoil approximation. We note that macroscopic quantum electrodynamics \cite{B12_2} provides an alternative framework to derive Eqs.~(\ref{chiPfinal}).

We note that Eq.~(\ref{GammaRR}), which relates the position-dependent electron decoherence to the electromagnetic Green tensor in the structure under consideration, bears some similarity to the electron energy-loss probability (EELS) \cite{paper149}. Within the linear and nonrecoil approximations, the latter can be derived by separately considering each frequency component of the external current $\jb(\rb,\omega)=-e\zz\,\ee^{\ii\omega z/v}\,\delta(\Rb-\Rb')$ associated with a classical electron beam passing by the transverse position $\Rb'$. The electric field generated by this current at a position $\rb$ is then obtained by applying the Green tensor as defined in Eq.~(\ref{Greentensor}), yielding 
\begin{align}
E_z(\rb,\Rb',\omega)=4\pi\ii\,e\,\omega \int_{-\infty}^\infty dz'\; G_{zz}(\rb,\rb',\omega)\;\ee^{\ii\omega z'/v}.
\label{practEz}
\end{align}
Following the methods in Ref.~\cite{paper149}, we find
\begin{align}
\tilde{\Gamma}(\Rb,\Rb',\omega)
&=\frac{e}{\pi\hbar\omega}\int_{-\infty}^\infty dz\;\Ree\big\{\ee^{-\ii\omega z/v}\,E_z(\rb,\Rb',\omega)\big\}
\label{practGtil} \\
&=\frac{4e^2}{\hbar}\int_{-\infty}^\infty dz\int_{-\infty}^\infty dz'\;\Imm\big\{-\ee^{-\ii\omega(z-z')/v}\, G_{zz}(\rb,\rb',\omega)\big\}
\nonumber
\end{align}
for a generalized EELS probability associated with a probe electron passing by the transverse position $\Rb$. In particular, $\Gamma(\Rb,\Rb,\omega)$ is the regular EELS probability. Finally, Eq.~(\ref{GammaRR}) can be recast as
\begin{align}
\Gamma(\Rb,\Rb',\omega)=\frac{1}{2}\big[\tilde{\Gamma}(\Rb',\Rb,\omega)+\tilde{\Gamma}(\Rb,\Rb',\omega)\big]
\label{practGamma}
\end{align}
when considering positions $z=z'$ (i.e., within the same plane in an interferometric measurement). In practice, we can then evaluate $P(\Rb,\Rb')$ in Eq.~(\ref{Prrfinal}) [with $z=z'$, so the result is independent of $z$ and $z'$ according to Eq.~(\ref{GammaRR})] by obtaining $\Gamma(\Rb,\Rb',\omega)$ from Eqs.~(\ref{practGtil}) and (\ref{practGamma}) using the frequency-space classical electric field $E_z(\rb,\Rb',\omega)$ [Eq.~(\ref{practEz})] produced at $\rb$ by an electron moving with velocity $\vb=v\,\zz$ and transverse coordinates $\Rb'$. It is useful to note that only the induced part of the electric field $\Eb^{\rm ind}$ makes a contribution because the direct field, related to the free-space Green tensor $G^0(\rb,\rb',\omega)$ [i.e., separating $G(\rb,\rb',\omega)=G^0(\rb,\rb',\omega)+G^{\rm ind}(\rb,\rb',\omega)$], only contains components proportional to $\ee^{\ii k_zz'}$ with a $z$-projected wave vector $|k_z|\le\omega/c$ (i.e., within the light cone), thus vanishing after integration in Eq.~(\ref{practEz}).

\renewcommand{\thesection}{C} 
\renewcommand{\theequation}{C\arabic{equation}} 
\section{Decoherence of an electron beam perpendicular to a perfectly conducting half-plane}
\label{appendixhalfplane}

We apply the formalism developed in Sec.~\ref{secboson} to an electron beam moving perpendicularly to a perfectly conducting metallic half-plane and prepared in a superposition of two paths defined by the transverse coordinates $\Rb_i$ with $i=1,2$ (see Fig.~2a in the main text). The half-plane is taken to occupy the $x<0$ region of the $z=0$ plane and the electron is moving with velocity $v$ along $z$ without intersecting the metal (i.e., $x_1,\,x_2>0$). We proceed by first calculating the induced electric field component $E_z^{\rm ind}(\rb,z,\Rb_i,\omega)$ produced by each of the paths $i$ [see Eq.~(\ref{practEz})], from which the decoherence probability $P(\Rb,\Rb')$ (for $z=z'$) is computed by using Eq.~(\ref{Prrfinal})  with $\Gamma(\Rb_i,z,\Rb_{i'},\omega)$ given by Eqs.~(\ref{practGtil}) and (\ref{practGamma}). As we argue at the end of Sec.~\ref{secboson}, the direct electric field produces a vanishing contribution to the decoherence probability, and therefore, we only calculate the field induced by the presence of the half-plane. It should also be noted that the present calculation can directly be applied to a laterally extended beam, but we retain the two-path terminology for clarity.

\subsection{Electron-induced electric field}

From the well-known expression for the electric field produced by an external source in free space \cite{J99}, the two-dimensional current $\jb_{i}^{\rm ind}(\Rb,\omega)$ (i.e., $\perp\zz$) induced on the half-plane by the electron path passing by $\Rb_{i}$ gives
\begin{align}
&\Eb^{\rm ind}(\rb,\Rb_{i},\omega) =\frac{\ii}{\omega} \big(k^2\,\mathcal{I}_3+\nabla \otimes \nabla\big)\cdot \int d^2\Rb'\; \frac{\ee^{\ii k |\rb-\Rb'|}}{|\rb-\Rb'|}\;\jb_{i}^{\rm ind}(\Rb',\omega),
\label{indField}
\end{align}
where $k=\omega/c$ is the light wavenumber, $\mathcal{I}_n$ denotes the $n\times n$ identity matrix, and we indicate the dependence on $\Rb_{i}$ by adopting the notation introduced in Eq.~(\ref{practEz}). Using the identity
\begin{align}
\frac{\ee^{\ii k |\rb-\Rb'|}}{|\rb-\Rb'|} = \frac{\ii}{2\pi} \int\frac{d^2\kparb}{k_z}\,\ee^{\ii\kparb\cdot (\Rb-\Rb')+\ii k_z|z|},
\label{kernel1}
\end{align}
with $\kparb=(k_x,k_y)$, $k_z=\sqrt{k^2-\kpar^2+\ii0^+}$, and $\Imm\{k_z\}>0$, we can work out Eq.~(\ref{indField}) to write the in-plane induced electric field as
\begin{align}
&\Eb^{\rm ind}_{\parallel}(\Rb,z=0,\Rb_{i},\omega) = -\frac{1}{2\pi\omega} \int\frac{d^2\kparb}{k_z}\,
\ee^{\ii\kparb\cdot\Rb} \mathcal{M}(\kparb,\omega)\cdot\jb_{i}^{\rm ind}(\kparb,\omega),
\label{inducedVec}
\end{align}
where $\jb_{i}^{\rm ind}(\kparb,\omega)=\int d^2\Rb'\; \ee^{-\ii\kparb\cdot\Rb}\;\jb_{i}^{\rm ind}(\Rb,\omega)$ is the induced current in momentum space and we define the $2\times2$ matrix $\mathcal{M}(\kparb,\omega)=k^2\mathcal{I}_2-\kparb\otimes\kparb$.

In the limit of a perfect conductor, both $\Eb_\parallel$ and the normal magnetic field $H_z$ must vanish on the surface of the half-plane. From Faraday's law, the vanishing of $\Eb_\parallel$ directly implies $H_z=(\ii/k)\zz\cdot(\nabla\times\Eb_\parallel)=(\ii/k)(\zz\times\nabla_\parallel)\cdot\Eb_\parallel=0$, so we only need to consider the electric field. We then have $\Eb^{\rm ind}_\parallel(\Rb,z=0,\Rb_i,\omega)=-\Eb^{\rm ext}_\parallel(\Rb,z=0,\Rb_{i},\omega)$, where \cite{paper149}
\begin{align}
\Eb_\parallel^{\rm ext}(\Rb,z=0,\Rb_{i},\omega)
&=\frac{\ii e}{\pi v}\int d^2\kparb\,\frac{\kparb\,\ee^{\ii\kparb\cdot(\Rb-\Rb_{i})}}{\kpar^2+(\omega/v\gamma)^2} \nonumber\\
&=\frac{e}{v} \int_{-\infty}^\infty\!\!\!dk_y\;\ee^{(\kappa,\ii k_y)\cdot(\Rb-\Rb_{i})}\;\big[\xx+(\ii k_y/\kappa)\,\yy\big]
\label{external}
\end{align}
is the in-plane external field due to the electron path passing by $\Rb_{i}$, $\gamma=1/\sqrt{1-v^2/c^2}$ is the relativistic Lorentz factor, and we define
\[\kappa=\sqrt{(\omega/v\gamma)^2+k_y^2}.\]
In Eq.~(\ref{external}), we obtain the rightmost expression by closing the integration contour over the lower complex $k_x$ plane (because we evaluate the field at positions $x<0<x_{i}$ in the half-plane), where only the pole $k_x=-\ii\kappa$ contributes. Combining Eqs.~(\ref{inducedVec}) and (\ref{external}), and working in $(x,k_y,\omega)$ space, the vanishing of the in-plane electric field at the half-plane leads to the condition
\begin{subequations}
\label{jconds}
\begin{align}
\int_{-\infty}^\infty\frac{dk_x}{k_z}\,
\ee^{\ii k_x x} \mathcal{M}(\kparb,\omega)\cdot\jb_{i}^{\rm ind}(\kparb,\omega)
=\eb_{i}(k_y,\omega)\;\ee^{\kappa x},
\label{jconds1}
\end{align}
for $x<0$, where
\[\eb_{i}(k_y,\omega)=\frac{2\pi\omega e}{v} \;\ee^{-(\kappa,\ii k_y)\cdot\Rb_{i}}\;\big[\xx+(\ii k_y/\kappa)\,\yy\big].\]
This needs to be supplemented by the condition of zero current outside the half-plane ($x>0$):
\begin{align}
\int_{-\infty}^\infty dk_x \;\ee^{\ii k_x x} \;\jb_{i}^{\rm ind}(\kparb,\omega)=0.
\label{jconds2}
\end{align}
\end{subequations}
The system formed by Eqs.~(\ref{jconds}) determines a unique solution for the current. Following the methods in Ref.~\cite{BW99}, we write
\begin{align}
k_z=\sqrt{K-k_x}\sqrt{K+k_x}  \label{kzkk}
\end{align}
with $K=\sqrt{k^2-k_y^2+\ii0^+}$ defined with the square root yielding a positive imaginary part. Noticing that the right-hand side of Eq.~(\ref{jconds1}) depends on $x$ just through a factor $\ee^{\kappa x}$, we can anticipate the solution
\begin{align}
\jb_{i}^{\rm ind}(\kparb,\omega)=\frac{\ii \omega e}{v K^2}
\frac{\ee^{-(\kappa,\ii k_y)\cdot\Rb_{i}}}{k_x+\ii\kappa}
\bigg[\frac{K^2\xx+k_xk_y\yy}{\sqrt{K+\ii\kappa}\sqrt{K+k_x}}+\frac{\ii k_y}{\kappa}\,\sqrt{K+\ii\kappa}\sqrt{K+k_x}\,\yy\bigg],
\label{jksol}
\end{align}
where the prefactor to the left of the square brackets incorporates a pole at $k_x=-\ii\kappa$, which gives rise to the aforementioned $x$ dependence upon integration of Eq.~(\ref{jconds1}). This expression is the superposition of the two possible solutions that eliminate the branch cut in the upper $k_x$ plane by either multiplying or dividing by $\sqrt{K+k_x}$, without introducing additional poles. The $\omega$- and $k_y$-dependent coefficients weighting these two solutions are determined by enforcing Eq.~(\ref{jconds1}), together with the condition that the electromagnetic energy remains finite despite the edge divergence \cite{J1950}, implying that the $x$ and $y$ components of the current must decay at least as $k_x^{-3/2}$ and $k_x^{-1/2}$, respectively, in the large $|k_x|$ limit. To verify that Eq.~(\ref{jksol}) is indeed a solution of Eq.~(\ref{jconds2}), we note that the integrand vanishes as $\Imm\{k_x\}\to\infty$, so we can close the integration contour over the upper complex $k_x$ plane. Although $\jb_{i}^{\rm ind}(\kparb,\omega)$ has one pole ($k_x=-\ii\kappa$) and one branch cut ($k_x=-K$), they are both lying on the lower plane, and consequently, the integral is indeed zero. Likewise, the integrand in Eq.~(\ref{jconds1}) vanishes as $\Imm\{k_x\}\to-\infty$ and has a single branch cut \cite{notedecoherence4} ($k_x=K$), which lies on the upper complex $k_x$ plane, so we close the integration contour over the lower plane, yielding the right-hand side of the equation via the $k_x=-\ii\kappa$ pole.

To study the interaction between the two electron paths, we need to consider the $z$ component of the field induced at $x>0$ [see Eq.~(\ref{practGtil})]. From Eq.~(\ref{indField}), using Eq.~(\ref{kernel1}) again, we find
\begin{align}
E^{\rm ind}_z(\rb,\Rb_{i},\omega)
&= \frac{{\rm sign}\{z\}}{2\pi\omega} \int d^2\kparb\,
\ee^{\ii(\kparb\cdot\Rb+k_z|z|)}\,\kparb\cdot\jb_{i}^{\rm ind}(\kparb,\omega),
\label{Eindfinal}
\end{align}
where $\jb_{i}^{\rm ind}(\kparb,\omega)$ is explicitly given by Eq.~(\ref{jksol}).

\subsection{Decoherence probability}

We obtain the generalized EELS probability by inserting Eq.~(\ref{Eindfinal}) into Eq.~(\ref{practGtil}) and carrying out the $z$ integral as
\begin{align}
\int_{-\infty}^\infty dz\,\ee^{\ii k_z|z|-\ii\omega z/v}\,{\rm sign}\{z\}=-\frac{2\ii\omega/v}{k_x^2+\kappa^2}.
\label{zintegral}
\end{align}
Noticing that $x_i>0$ at the path positions $i=1,2$, we can perform the $k_x$ integral by closing the contour over the upper complex plane, where $\jb_{i}^{\rm ind}(\kparb,\omega)$ is free from poles and branch cuts, while the result in Eq.~(\ref{zintegral}) contributes with a $k_x=\ii\kappa$ pole, yielding
\begin{align}
\tilde{\Gamma}(\Rb_1,\Rb_2,\omega)=&\frac{e^2}{\pi\hbar c^2} \;
\int_0^k
\frac{dk_y}{\kappa^3}
\;\frac{\kappa^2+(c^2/v^2) k_y^2}{\sqrt{k^2-k_y^2}} \; \ee^{-\kappa (x_1+x_2)}
\,\cos\big[k_y(y_1-y_2)\big].
\nonumber
\end{align}
In this particular instance, we find $\tilde{\Gamma}(\Rb_1,\Rb_2,\omega)=\tilde{\Gamma}(\Rb_2,\Rb_1,\omega)$, so Eq.~(\ref{practGamma}) becomes
\begin{align}
\Gamma(\Rb_1,\Rb_2,\omega) = &\frac{\alpha}{\pi\omega}
\int_0^1 \frac{d\mu}{\sqrt{1-\mu^2}}\;\frac{\big[(1+v^2/c^2)\mu^2+\eta^2\big]}{\big(\mu^2+\eta^2\big)^{3/2}}
\; \cos\big[\mu\,k(y_1-y_2)\big]
\;\ee^{-k(x_1+x_2)\sqrt{\mu^2+\eta^2}},
\label{GRRw}
\end{align}
where $\alpha = e^2/\hbar c\approx1/137$ is the fine structure constant, $\eta=c/v\gamma$ is a velocity-dependent parameter, and $\mu =k_y/k$. Finally, we insert Eq.~(\ref{GRRw}) into Eq.~(\ref{Prrfinal}) to obtain the decoherence probability
\begin{align}
P(\Rb_1,\Rb_2)=&\frac{\alpha}{2\pi}
\int_0^1 \frac{d\mu}{\sqrt{1-\mu^2}}\;\frac{\big[(1+v^2/c^2)\mu^2+\eta^2\big]}{\big(\mu^2+\eta^2\big)^{3/2}}
\int_0^\infty \frac{d\theta}{\theta} \;\coth(\theta/4\pi)  \nonumber\\
&\times \bigg[ \ee^{-2\theta\,(d_1/\lambda_T)\sqrt{\mu^2+\eta^2}} + \ee^{-2\theta\,(d_2/\lambda_T)\sqrt{\mu^2+\eta^2}}
-2\cos\big(\mu\theta\,d_\perp/\lambda_T\big)\,\ee^{-\theta\,[(d_1+d_2)/\lambda_T]\sqrt{\mu^2+\eta^2}}\bigg] \nonumber 
\end{align}
[Eq.~(4) in the main text], where we have made the changes of variables
$\theta=2\pi\hbar\omega/\kB T$ and $\mu =k_y/k$.

\renewcommand{\thesection}{D} 
\renewcommand{\theequation}{D\arabic{equation}} 
\section{Decoherence by interaction with a metallic ribbon}
\label{ribbon}

To investigate the role played by the size of the structure, we consider a zero-thickness, perfectly conducting ribbon of width $W$ lying on the $z=0$ plane, defined by edges at $x=0$ and $x=-W$, and having infinite extension along $y$. We start from Eq.~(\ref{jconds1}), now restricted to the $-W<x<0$ region (the ribbon) and write $\jb_{i}^{\rm ind}(\kparb,\omega) = \int_{-W}^0 dx'\, e^{-\ii k_x x'}\,\jb_{i}^{\rm ind}(x',k_y,\omega)$ in terms of the $x$-space current $\jb_{i}^{\rm ind}(x,k_y,\omega)$ (also limited to the ribbon area). After inserting this expression into Eq.~(\ref{jconds1}), we carry out the $k_x$ integral by using the identity $\int_{-\infty}^\infty dk_x\,\ee^{\ii k_x (x-x')}\,\big(k^2-k_y^2-k_x^2+\ii0^+\big)^{-1/2}=2\ii K_0\big(Q\,|x-x'|\big)$ (see Eq.~3.754-2 in Ref.~\cite{GR1980}), where we define
\[Q=\sqrt{k_y^2-k^2-\ii0^+}\]
with the square root taken to yield a positive real part. This leads to
\begin{align}
2\ii \int_{-W}^{0} dx'\;\big[k^2\mathcal{I}_2+(\partial_x,\ii k_y)\otimes(\partial_x,\ii k_y)\big]\cdot\jb_{i}^{\rm ind}(x',k_y,\omega)
\;K_0\big(Q\,|x-x'|\big)
=\eb_{i}(k_y,\omega)\,\ee^{\kappa x}
\label{condRib1}
\end{align}
with $x$ in the $(-W,0)$ range. Incidentally, this expression can also be obtained starting from the induced field in Eq.~(\ref{indField}) after Fourier-transforming in $y$.

We find $\jb_{i}^{\rm ind}(x,k_y,\omega)$ numerically from Eq.~(\ref{condRib1}) by discretizing the integral through a set of $N$ equally spaced points $\tilde{x}_j=-W+(j+1/2)W/N$ labeled by $j=0,\cdots,N-1$, each of them representing an interval $\tilde{x}_j-h/2<x<\tilde{x}_j+h/2$ of width $h=W/N$. By approximating
\begin{align}
\jb_{i}^{\rm ind}(x,k_y,\omega)\approx\jb_{i}^{\rm ind}(\tilde{x}_j,k_y,\omega)\equiv(2\pi\omega e/v)\;\ee^{-(\kappa,\ii k_y)\cdot\Rb_{i}}\;\Jb_j(k_y,\omega)
\label{Japprox}
\end{align}
(i.e., constant within each interval $j$), we transform Eq.~(\ref{condRib1}) into the $N\times N$ linear system
\begin{align}
\sum_{j'=0}^{N-1}
\Big[\big(k^2\mathcal{I}_2-k_y^2\,\yy\otimes\yy\big)\;I^0_{jj'}
+\ii k_y\,\big(\xx\otimes\yy+\yy\otimes\xx\big)\;I^1_{jj'}
+\xx\otimes\xx\;I^2_{jj'}\Big]\cdot\Jb_{j'}(k_y,\omega)=\big[\xx+(\ii k_y/\kappa)\,\yy\big]\,\ee^{\kappa \tilde{x}_j},
\nonumber
\end{align}
where $I^n_{jj'}=H^n(\tilde{x}_j-\tilde{x}_{j'}+h/2)-H^n(\tilde{x}_j-\tilde{x}_{j'}-h/2)$ are interval integrals defined in terms of the functions
\begin{align}
H^0(x)&=2\ii\int dx\;K_0\big(Q\,|x|\big)
=-\ii\,\pi\, x\, \Big[\textbf{L}_{-1}\big(Q\,|x|\big)\,K_0\big(Q\,|x|\big)+\textbf{L}_{0}\big(Q\,|x|\big)\,K_1\big(Q\,|x|\big)\Big]+C, \nonumber\\
H^1(x)&=2\ii\int dx\;\partial_x K_0\big(Q\,|x|\big) =-2\,\ii\;K_0\big(Q\,|x|\big)+C, \nonumber\\
H^2(x)&=2\ii\int dx\;\partial_{xx}K_0\big(Q\,|x|\big) =2\,\ii\;{\rm sig}\{x\}\; Q\,K_1\big(Q\,|x|\big)+C,
\nonumber
\end{align}
with $\textbf{L}_n$ denoting modified Struve functions.

From the induced current, we then combine Eqs.~(\ref{practGtil}), (\ref{Eindfinal}), and (\ref{zintegral}) to obtain
\begin{align}
\tilde{\Gamma}(\Rb_1,\Rb_2,\omega) &= \frac{e}{\pi\hbar\omega v} \int_{-\infty}^\infty dk_y \int_{-W}^0 dx\;
\ee^{\kappa x}
\;\Ree\Big\{\ee^{(-\kappa,\ii k_y)\cdot\Rb_1}\;\big[\xx-\ii(k_y/\kappa)\yy\big]\cdot\jb_2^{\rm ind}(x,k_y,\omega)\Big\}.
\nonumber
\end{align}
Then, using the discretized solution [see Eq.~(\ref{Japprox})] and applying Eq.~(\ref{practGamma}), we find
\begin{align}
\Gamma(\Rb_1,\Rb_2,\omega) =\frac{4\alpha c}{v^2} &\int_{-\infty}^\infty \frac{dk_y}{\kappa}
\,\sinh(\kappa h/2)\,\ee^{-\kappa(x_1+x_2)}\,\cos\big[k_y(y_1-y_2)\big] \; \Ree\bigg\{\big[\xx-\ii(k_y/\kappa)\yy\big]\cdot\sum_{j=0}^{N-1} \ee^{\kappa \tilde{x}_j}\Jb_j(k_y,\omega) \bigg\}.
\nonumber
\end{align}
Finally, we compute the decoherence probability by inserting this expression into Eq.~(\ref{Prrfinal}).

\renewcommand{\thesection}{E} 
\renewcommand{\theequation}{E\arabic{equation}} 
\section{Low-frequency divergence of the EELS probability in extended structures}
\label{divergencesinGamma}

In the introduction of the main text, we argue that the frequency-resolved EELS probability $\Gamma(\omega)$ (i.e., a function of energy loss $\hbar\omega$) vanishes in the low-frequency limit for finite objects, but it diverges for extended structures. At zero temperature, the EELS probability is given by Eq.~(\ref{GammaRR}) with $\rb=\rb'$. For a finite particle and small $\omega$, this expression is proportional to ${\rm Im}\{\alpha(\omega)\}$ \cite{paper371}, where $\alpha(\omega)$ is the polarizability. In the presence of inelastic absorption, we have ${\rm Im}\{\alpha(\omega)\}\propto\omega$, while in the absence of absorption the optical theorem \cite{V1981} (i.e., ${\rm Im}\{\alpha^{-1}(\omega)\}=2k^3/3$) prescribes a probability $\propto\omega^3$ due to radiative losses. This result can be intuitively understood from the increasingly small size of the particle compared to the light wavelength as $\omega$ is reduced. In addition, at finite temperature $T$, the inelastic scattering probability associated with the coupling to bosonic modes (e.g., photons and polaritons) must be multiplied by a factor $n_T(\omega)+1$ for electron energy losses ($\omega>0$) and $n_T(-\omega)$ for gains ($\omega<0$) \cite{paper371}, where the Bose-Einstein distribution function $n_T(\omega)$ [Eq.~(\ref{nTw})] introduces an additional factor $\propto1/\omega$ at low frequencies. Consequently, the low-$\omega$ scaling of the loss probability for finite structures at finite temperatures is $\propto\omega^0$ in the presence of inelastic absorption and $\propto\omega^2$ for lossless objects. These conclusions are summarized in Table~1 in the main text (left columns).

For extended structures (e.g., a half-plane), the EELS probability was shown to diverge as $\propto1/\omega$ at zero temperature due to radiative coupling \cite{paper127}. The divergence becomes more dramatic (as $\propto1/\omega^2$) at finite temperature due to the additional factor $n_T(\omega)+1$ for energy losses and $n_T(\omega)$ for gains. These conclusions are summarized in Table~1 in the main text (right column). Reassuringly, these divergences do not produce any unphysical results, and in particular, the frequency-integrated electron energy change [i.e., $\Delta E=\int_{-\infty}^\infty d\omega\,\hbar\omega\,\Gamma(\omega)$] is temperature-independent because the $n_T(\omega)$ terms cancel when adding losses and gains. Furthermore, $\Delta E$ is finite because the remaining $\propto1/\omega$ divergence is suppressed when multiplied by the energy loss $\hbar\omega$. We also note that the decoherence probability studied above involves a similar cancelation among $\Gamma(\Rb,\Rb',\omega)$ terms in Eq.~(2) of the main text.

\renewcommand{\thesection}{F} 
\renewcommand{\theequation}{F\arabic{equation}} 
\section{Fluctuation-dissipation theorem for the vector potential}
\label{AppendixA}

We intend to derive Eqs.~(\ref{AAaverages1}) and (\ref{AAaverages2}), which are essentially the fluctuation-dissipation theorem \cite{N1928,CW1951} for the vector potential operator $\Abh(\rb,t)$. Our starting point is an interaction Hamiltonian
\begin{align}
\HH_{\rm int}(t)=-\frac{1}{c}\int d^3\rb\;\jb(\rb,t)\cdot\Abh(\rb,t) \nonumber
\end{align}
in the interaction picture within the so-called temporal gauge (i.e., with the scalar potential set to zero), with $\jb(\rb,t)$ representing a classical external current. We also consider a complete set of states $\ket{n}$ of the free radiation Hamiltonian $\HH_{\rm rad}$, satisfying $\HH_{\rm rad}\ket{n}=\hbar\omega_n\ket{n}$. Applying first-order perturbation theory under the assumption of a vanishing interaction in the infinite past [i.e., $\HH_{\rm int}(-\infty)=0]$, the time-dependent perturbed states become
\begin{align}
\ket{\psi_n(t)}\approx\ket{n}-\frac{\ii}{\hbar}\int_{-\infty}^t dt'\,\HH_{\rm int}(t')\ket{n}. \nonumber
\end{align}
We now introduce a thermal average at temperature $T$, so the expected value of the induced vector potential reduces to
\begin{align}
A_a(\rb,t)=\braket{\Ah_a(\rb,t)}_T
&=\frac{1}{Z}\sum_n\ee^{-\hbar\omega_n/\kB T}\big[\bra{\psi_n(t)}\Ah_a(\rb,t)\ket{\psi_n(t)}-\bra{n}\Ah_a(\rb,t)\ket{n}\big] \nonumber\\
&\approx\frac{\ii}{\hbar c\,Z}\sum_{n,a'} \ee^{-\hbar\omega_n/\kB T}\int^t_{-\infty}dt'\int d^3\rb' \bra{n}\big[\Ah_a(\rb,t),\Ah_{a'}(\rb',t')\big]\ket{n} \,j_{a'}(\rb',t') \nonumber\\
&=-4\pi c \int d^3\rb'\int_{-\infty}^\infty dt'\sum_{a'}\;G_{aa'}(\rb, \rb', t-t') \,j_{a'}(\rb',t') \nonumber
\end{align}
to first order in the external current. Here, $Z=\sum_n\ee^{-\hbar\omega_n/\kB T}$ is the partition function,
\begin{align}
&G_{aa'}(\rb, \rb',t-t')=-\frac{\ii}{4\pi\hbar c^2\,Z}\Theta(t-t') \; \sum_n \ee^{-\hbar\omega_n/\kB T} \bra{n}\big[\Ah_a(\rb,t),\Ah_{a'}(\rb',t')\big]\ket{n}
\label{Greendefinition}
\end{align}
is the electromagnetic Green tensor, and we use indices $a$ and $a'$ to label Cartesian components. The Green tensor must satisfy Eq.~(\ref{Greentensor}) to guarantee that the average potential $\Ab(\rb,t)$ fulfills the macroscopic Maxwell equations in the presence of a source current $\jb(\rb,t)$ \cite{AGD1965}. We move to the frequency domain by Fourier-transforming the Green tensor as
\begin{align}
G_{aa'}(\rb, \rb', \omega)&=\int_{-\infty}^\infty d\tau\; G_{aa'}(\rb, \rb',\tau)\,\ee^{\ii\omega\tau} \nonumber\\
&=\frac{1}{4\pi\hbar c^2\,Z} \sum_{nn'} \bra{n}\Ah_a(\rb)\ket{n'}\bra{n'}\Ah_{a'}(\rb')\ket{n} \;\frac{\ee^{-\hbar\omega_n/\kB T}-\ee^{-\hbar\omega_{n'}/\kB T}}{\omega+\omega_n-\omega_{n'}+\ii0^+} \nonumber\\
&=\frac{1}{4\pi\hbar c^2} \int_{-\infty}^\infty \frac{d\omega'}{\omega-\omega'+\ii0^+}\;\frac{S_{aa'}(\rb,\rb',\omega')}{n_T(\omega')+1}, 
\label{GFDT}
\end{align}
where the second line is obtained by inserting the unity operator $\sum_n\ket{n}\bra{n}$ in the $\Ah\Ah$ products of Eq.~(\ref{Greendefinition}) and further writing $\Abh(\rb,t)= \ee^{\ii\HH_{\rm rad}t/\hbar}\Abh(\rb)\ee^{-\ii\HH_{\rm rad}t/\hbar}$ in terms of the vector potential operator $\Abh(\rb)$ in the Schr\"odinger picture. Then, the third line is expressed in terms of the Bose-Einstein distribution function $n_T(\omega)$ [Eq.~(\ref{nTw})] and we define the spectral tensor
\begin{align}
&S_{aa'}(\rb,\rb',\omega)
=\frac{1}{Z}\sum_{nn'}\ee^{-\hbar\omega_n/k_B T} \; \bra{n}\Ah_a(\rb)\ket{n'}\bra{n'}\Ah_{a'}(\rb')\ket{n} \,\delta(\omega+\omega_n-\omega_{n'}).
\nonumber
\end{align}
By construction, the Green tensor in Eq.~(\ref{GFDT}) trivially satisfies the causality property
\begin{subequations}
\label{causalityreciprocity}
\begin{align}
G(\rb,\rb',\omega)=G^*(\rb,\rb',-\omega),
\label{causality}
\end{align}
and its poles are all lying on the lower complex $\omega$ plane. In addition, the spectral tensor is real for systems with time-reversal symmetry \cite{BT1962},
implying that the reciprocity property $S_{aa'}(\rb,\rb',\omega)=S_{a'a}(\rb',\rb,\omega)$ is fulfilled and further transferred to the Green tensor:
\begin{align}
G_{aa'}(\rb,\rb',\omega)=G_{a'a}(\rb',\rb,\omega).
\label{reciprocity}
\end{align}
\end{subequations}
Since the spectral tensor is real, we can write it as
\begin{align}
S_{aa'}(\rb,\rb',\omega)=-4\hbar c^2\;[n_T(\omega)+1]\;\Imm\big\{G_{aa'}(\rb, \rb', \omega)\big\}
\label{SnG}
\end{align}
directly from Eq.~(\ref{GFDT}). We note that a temperature dependence is introduced in the Green tensor through the thermal average in Eq.~(\ref{Greendefinition}), which we dismiss in this work because it is small unless high temperatures affecting the properties of the involved materials are considered.

We follow an analogous procedure to calculate the correlations of the vector potential operator in the frequency domain, $\Abh(\rb,\omega)=\int dt\; \ee^{\ii\omega t}\,\Abh(\rb,t)$. Using the unity operator $\sum_n\ket{n}\bra{n}$ and expressing the potentials in the Schr\"odinger picture, one finds
\begin{align}
\big\langle\Ah_a(\rb,\omega)\Ah_{a'}(\rb',\omega')\big\rangle_T
&=\frac1{Z}\int_{-\infty}^\infty dt\int_{-\infty}^\infty dt'\; \ee^{\ii\omega t+\ii\omega't'}\sum_{nn'}\ee^{-\hbar\omega_n/k_B T} \; \ee^{i(\omega_n-\omega_{n'})(t-t')} \bra{n}\Ah_a(\rb)\ket{n'}\bra{n'}\Ah_{a'}(\rb')\ket{n} \nonumber \\
&=4\pi^2\;S_{aa'}(\rb,\rb',\omega)\;\delta(\omega+\omega'), \nonumber
\end{align}
which, together with Eq.~(\ref{SnG}), leads to
\begin{align}
&\big\langle\Ah_a(\rb,\omega)\Ah_{a'}(\rb',\omega')\big\rangle_T =-16\pi^2\,\hbar c^2\; \delta(\omega+\omega') \; [n_T(\omega)+1]\;\Imm\big\{G_{aa'}(\rb,\rb',\omega)\big\}.
\label{AAwaverage}
\end{align}
Finally, performing the inverse Fourier transform of Eq.~(\ref{AAwaverage}) (in both $\omega$ and $\omega'$) and using causality [Eq.~(\ref{causality})] together with the identity $n_T(\omega)+1=-n_T(-\omega)$, we readily obtain Eq.~(\ref{AAaverages1}).

Now, exchanging the order of the $\Ah$ operators in Eq.~(\ref{AAwaverage}) and applying reciprocity and causality [Eqs.~(\ref{causalityreciprocity})], we obtain
\begin{align}
&\big\langle\Ah_{a'}(\rb',\omega')\Ah_a(\rb,\omega)\big\rangle_T =-16\pi^2\,\hbar c^2\; \delta(\omega+\omega')\; n_T(\omega)\;\Imm\big\{G_{aa'}(\rb,\rb',\omega)\big\},
\nonumber
\end{align}
which, combined with Eq.~(\ref{AAwaverage}), produces the commutator
\begin{align}
&\big\langle[\Ah_a(\rb,\omega),\Ah_{a'}(\rb',\omega')]\big\rangle_T =-16\pi^2\,\hbar c^2\; \delta(\omega+\omega') \; \Imm\big\{G_{aa'}(\rb,\rb',\omega)\big\},
\nonumber
\end{align}
whose inverse Fourier transform directly yields Eq.~(\ref{AAaverages2}). 

\renewcommand{\thesection}{G} 
\renewcommand{\theequation}{G\arabic{equation}} 
\section{Derivation of Eq.~(\ref{GGfinal})}
\label{AppendixB}

We use the commutator in Eq.~(\ref{AAaverages2}) to transform Eq.~(\ref{Gop}) into
\begin{align}
\Gop(\rb)=&\ii\,\frac{8e^2v^2}{\hbar}
\int_{-\infty}^\infty dt\int_{-\infty}^t dt'\int_0^\infty d\omega\;
\;\sin\big[\omega(t-t')\big] \; \Imm\big\{G_{zz}(\rb-\vb t,\rb-\vb t',\omega)\big\}. \nonumber
\end{align}
Inverting the order of the $t$ and $t'$ integrals, exchanging the $t$ and $t'$ variables, and using reciprocity [Eq.~(\ref{reciprocity})], we find
\begin{align}
\Gop(\rb)=&-\ii\,\frac{8e^2v^2}{\hbar}
\int_{-\infty}^\infty dt\int_t^\infty dt'\int_0^\infty d\omega\;
\;\sin\big[\omega(t-t')\big] \; \Imm\big\{G_{zz}(\rb-\vb t,\rb-\vb t',\omega)\big\}. \nonumber
\end{align}
The average of these two expressions produces the result
\begin{align}
\Gop(\rb)=&\ii\,\frac{4e^2v^2}{\hbar}
\int_{-\infty}^\infty dt\int_{-\infty}^\infty dt'\int_0^\infty d\omega\;
\;\sin\big(\omega|t-t'|\big) \; \Imm\big\{G_{zz}(\rb-\vb t,\rb-\vb t',\omega)\big\}, \label{N2inter}
\end{align}
where the argument of the sine function involves the absolute value of the time difference. Now, we consider the identity \cite{notedecoherence5}
%
\begin{align}
&\int_0^\infty d\omega\;\sin\big(\omega|\tau|\big)\;\Imm\big\{\chi(\omega)\big\} =\int_0^\infty d\omega\;\cos(\omega\tau)\;\Ree\big\{\chi(\omega)\big\},
\label{sfsc}
\end{align}
which is valid for any response function satisfying $\chi(\omega)=\chi^*(-\omega)$ and having no poles with $\Imm\{\omega\}>0$. Applying Eq.~(\ref{sfsc}) to Eq.~(\ref{N2inter}) with $\tau=t-t'$ and $\chi(\omega)=G_{zz}(\rb-\vb t,\rb-\vb t',\omega)$, and changing the integration variables from $t$ and $t'$ to $z-vt$ and $z-vt'$, we finally obtain Eq.~(\ref{GGfinal}).
\end{widetext}


\begin{thebibliography}{68}
\expandafter\ifx\csname natexlab\endcsname\relax\def\natexlab#1{#1}\fi
\expandafter\ifx\csname bibnamefont\endcsname\relax
  \def\bibnamefont#1{#1}\fi
\expandafter\ifx\csname bibfnamefont\endcsname\relax
  \def\bibfnamefont#1{#1}\fi
\expandafter\ifx\csname citenamefont\endcsname\relax
  \def\citenamefont#1{#1}\fi
\expandafter\ifx\csname url\endcsname\relax
  \def\url#1{\texttt{#1}}\fi
\expandafter\ifx\csname urlprefix\endcsname\relax\def\urlprefix{URL }\fi
\providecommand{\bibinfo}[2]{#2}
\providecommand{\eprint}[2][]{\url{#2}}

\bibitem[{\citenamefont{Egerton}(2005)}]{E05}
\bibinfo{author}{\bibfnamefont{R.~F.} \bibnamefont{Egerton}},
  \emph{\bibinfo{title}{Physical Principles of Electron Microscopy: An
  Introduction to {TEM}, {SEM}, and {AEM}}} (\bibinfo{publisher}{Springer},
  \bibinfo{address}{New York}, \bibinfo{year}{2005}).

\bibitem[{\citenamefont{Muller et~al.}(2008)\citenamefont{Muller, {Fitting
  Kourkoutis}, Murfitt, Song, Hwang, Silcox, Dellby, and Krivanek}}]{MKM08}
\bibinfo{author}{\bibfnamefont{D.~A.} \bibnamefont{Muller}},
  \bibinfo{author}{\bibfnamefont{L.}~\bibnamefont{{Fitting Kourkoutis}}},
  \bibinfo{author}{\bibfnamefont{M.}~\bibnamefont{Murfitt}},
  \bibinfo{author}{\bibfnamefont{J.~H.} \bibnamefont{Song}},
  \bibinfo{author}{\bibfnamefont{H.~Y.} \bibnamefont{Hwang}},
  \bibinfo{author}{\bibfnamefont{J.}~\bibnamefont{Silcox}},
  \bibinfo{author}{\bibfnamefont{N.}~\bibnamefont{Dellby}}, \bibnamefont{and}
  \bibinfo{author}{\bibfnamefont{O.~L.} \bibnamefont{Krivanek}},
  \bibinfo{journal}{Science} \textbf{\bibinfo{volume}{319}},
  \bibinfo{pages}{1073} (\bibinfo{year}{2008}).

\bibitem[{\citenamefont{C.Davisson and Germer}(1927)}]{DG1927}
\bibinfo{author}{\bibnamefont{C.Davisson}} \bibnamefont{and}
  \bibinfo{author}{\bibfnamefont{L.~H.} \bibnamefont{Germer}},
  \bibinfo{journal}{Phys.\ Rev.} \textbf{\bibinfo{volume}{30}},
  \bibinfo{pages}{705} (\bibinfo{year}{1927}).

\bibitem[{\citenamefont{Pendry}(1974)}]{P1974}
\bibinfo{author}{\bibfnamefont{J.~B.} \bibnamefont{Pendry}},
  \emph{\bibinfo{title}{Low Energy Electron Diffraction}}
  (\bibinfo{publisher}{Academic Press}, \bibinfo{address}{London},
  \bibinfo{year}{1974}).

\bibitem[{\citenamefont{Rehr et~al.}(2002)\citenamefont{Rehr, Schattke,
  {Garc\'{\i}a de Abajo}, {D\'{\i}ez Mui\~{n}o}, and {Van Hove}}}]{paper037}
\bibinfo{author}{\bibfnamefont{J.~J.} \bibnamefont{Rehr}},
  \bibinfo{author}{\bibfnamefont{W.}~\bibnamefont{Schattke}},
  \bibinfo{author}{\bibfnamefont{F.~J.} \bibnamefont{{Garc\'{\i}a de Abajo}}},
  \bibinfo{author}{\bibfnamefont{R.}~\bibnamefont{{D\'{\i}ez Mui\~{n}o}}},
  \bibnamefont{and} \bibinfo{author}{\bibfnamefont{M.~A.} \bibnamefont{{Van
  Hove}}}, \bibinfo{journal}{J.\ Electron\ Spectrosc.\ Relat.\ Phenom.}
  \textbf{\bibinfo{volume}{126}}, \bibinfo{pages}{67} (\bibinfo{year}{2002}).

\bibitem[{\citenamefont{Wolter et~al.}(2016)\citenamefont{Wolter, Pullen, Le,
  Baudisch, Doblhoff-Dier, Senftleben, Hemmer, Schr\"oter, Ullrich, Pfeifer
  et~al.}}]{WPL16}
\bibinfo{author}{\bibfnamefont{B.}~\bibnamefont{Wolter}},
  \bibinfo{author}{\bibfnamefont{M.~G.} \bibnamefont{Pullen}},
  \bibinfo{author}{\bibfnamefont{A.-T.} \bibnamefont{Le}},
  \bibinfo{author}{\bibfnamefont{M.}~\bibnamefont{Baudisch}},
  \bibinfo{author}{\bibfnamefont{K.}~\bibnamefont{Doblhoff-Dier}},
  \bibinfo{author}{\bibfnamefont{A.}~\bibnamefont{Senftleben}},
  \bibinfo{author}{\bibfnamefont{M.}~\bibnamefont{Hemmer}},
  \bibinfo{author}{\bibfnamefont{C.~D.} \bibnamefont{Schr\"oter}},
  \bibinfo{author}{\bibfnamefont{J.}~\bibnamefont{Ullrich}},
  \bibinfo{author}{\bibfnamefont{T.}~\bibnamefont{Pfeifer}},
  \bibnamefont{et~al.}, \bibinfo{journal}{Science}
  \textbf{\bibinfo{volume}{354}}, \bibinfo{pages}{308} (\bibinfo{year}{2016}).

\bibitem[{\citenamefont{Filippetto et~al.}(2022)\citenamefont{Filippetto,
  Musumeci, Li, Siwick, Otto, Centurion, and Nunes}}]{FML22}
\bibinfo{author}{\bibfnamefont{D.}~\bibnamefont{Filippetto}},
  \bibinfo{author}{\bibfnamefont{P.}~\bibnamefont{Musumeci}},
  \bibinfo{author}{\bibfnamefont{R.~K.} \bibnamefont{Li}},
  \bibinfo{author}{\bibfnamefont{B.~J.} \bibnamefont{Siwick}},
  \bibinfo{author}{\bibfnamefont{M.~R.} \bibnamefont{Otto}},
  \bibinfo{author}{\bibfnamefont{M.}~\bibnamefont{Centurion}},
  \bibnamefont{and} \bibinfo{author}{\bibfnamefont{J.~P.~F.}
  \bibnamefont{Nunes}}, \bibinfo{journal}{Rev.\ Mod.\ Phys.}
  \textbf{\bibinfo{volume}{94}}, \bibinfo{pages}{045004}
  (\bibinfo{year}{2022}).

\bibitem[{\citenamefont{Anglin et~al.}(1997)\citenamefont{Anglin, Paz, and
  Zurek}}]{APZ97}
\bibinfo{author}{\bibfnamefont{J.~R.} \bibnamefont{Anglin}},
  \bibinfo{author}{\bibfnamefont{J.~P.} \bibnamefont{Paz}}, \bibnamefont{and}
  \bibinfo{author}{\bibfnamefont{W.~H.} \bibnamefont{Zurek}},
  \bibinfo{journal}{Phys.\ Rev.\ A} \textbf{\bibinfo{volume}{55}},
  \bibinfo{pages}{4041} (\bibinfo{year}{1997}).

\bibitem[{\citenamefont{Mazzitelli et~al.}(2003)\citenamefont{Mazzitelli, Paz,
  and Villanueva}}]{MPV03}
\bibinfo{author}{\bibfnamefont{F.~D.} \bibnamefont{Mazzitelli}},
  \bibinfo{author}{\bibfnamefont{J.~P.} \bibnamefont{Paz}}, \bibnamefont{and}
  \bibinfo{author}{\bibfnamefont{A.}~\bibnamefont{Villanueva}},
  \bibinfo{journal}{Phys.\ Rev.\ A} \textbf{\bibinfo{volume}{68}},
  \bibinfo{pages}{062106} (\bibinfo{year}{2003}).

\bibitem[{\citenamefont{Levinson}(2004)}]{L04}
\bibinfo{author}{\bibfnamefont{Y.}~\bibnamefont{Levinson}},
  \bibinfo{journal}{J.\ Phys.\ A:\ Math.\ Gen.} \textbf{\bibinfo{volume}{37}},
  \bibinfo{pages}{3003} (\bibinfo{year}{2004}).

\bibitem[{\citenamefont{Hsiang and Lee}(2006)}]{HL06}
\bibinfo{author}{\bibfnamefont{J.-T.} \bibnamefont{Hsiang}} \bibnamefont{and}
  \bibinfo{author}{\bibfnamefont{D.-S.} \bibnamefont{Lee}},
  \bibinfo{journal}{Phys.\ Rev.\ D} \textbf{\bibinfo{volume}{73}},
  \bibinfo{pages}{065022} (\bibinfo{year}{2006}).

\bibitem[{\citenamefont{Machnikowski}(2006)}]{M06_3}
\bibinfo{author}{\bibfnamefont{P.}~\bibnamefont{Machnikowski}},
  \bibinfo{journal}{Phys.\ Rev.\ B} \textbf{\bibinfo{volume}{73}},
  \bibinfo{pages}{155109} (\bibinfo{year}{2006}).

\bibitem[{\citenamefont{Howie}(2011)}]{H11_2}
\bibinfo{author}{\bibfnamefont{A.}~\bibnamefont{Howie}},
  \bibinfo{journal}{Ultramicroscopy} \textbf{\bibinfo{volume}{111}},
  \bibinfo{pages}{761} (\bibinfo{year}{2011}).

\bibitem[{\citenamefont{Scheel and Buhmann}(2012)}]{SB12}
\bibinfo{author}{\bibfnamefont{S.}~\bibnamefont{Scheel}} \bibnamefont{and}
  \bibinfo{author}{\bibfnamefont{S.~Y.} \bibnamefont{Buhmann}},
  \bibinfo{journal}{Phys.\ Rev.\ A} \textbf{\bibinfo{volume}{85}},
  \bibinfo{pages}{030101} (\bibinfo{year}{2012}).

\bibitem[{\citenamefont{Sonnentag and Hasselbach}(2007)}]{SH07}
\bibinfo{author}{\bibfnamefont{P.}~\bibnamefont{Sonnentag}} \bibnamefont{and}
  \bibinfo{author}{\bibfnamefont{F.}~\bibnamefont{Hasselbach}},
  \bibinfo{journal}{Phys.\ Rev.\ Lett.} \textbf{\bibinfo{volume}{98}},
  \bibinfo{pages}{200402} (\bibinfo{year}{2007}).

\bibitem[{\citenamefont{Hasselbach}(2010)}]{H10_2}
\bibinfo{author}{\bibfnamefont{F.}~\bibnamefont{Hasselbach}},
  \bibinfo{journal}{Rep.\ Prog.\ Phys.} \textbf{\bibinfo{volume}{73}},
  \bibinfo{pages}{016101} (\bibinfo{year}{2010}).

\bibitem[{\citenamefont{Beierle et~al.}(2018)\citenamefont{Beierle, Zhang, and
  Batelaan}}]{BZB18}
\bibinfo{author}{\bibfnamefont{P.~J.} \bibnamefont{Beierle}},
  \bibinfo{author}{\bibfnamefont{L.}~\bibnamefont{Zhang}}, \bibnamefont{and}
  \bibinfo{author}{\bibfnamefont{H.}~\bibnamefont{Batelaan}},
  \bibinfo{journal}{New\ J.\ Phys.} \textbf{\bibinfo{volume}{20}},
  \bibinfo{pages}{113030} (\bibinfo{year}{2018}).

\bibitem[{\citenamefont{Kerker et~al.}(2020)\citenamefont{Kerker, R\"{o}pke,
  Steinert, Pooch, and Stibor}}]{KRS20}
\bibinfo{author}{\bibfnamefont{N.}~\bibnamefont{Kerker}},
  \bibinfo{author}{\bibfnamefont{R.}~\bibnamefont{R\"{o}pke}},
  \bibinfo{author}{\bibfnamefont{L.~M.} \bibnamefont{Steinert}},
  \bibinfo{author}{\bibfnamefont{A.}~\bibnamefont{Pooch}}, \bibnamefont{and}
  \bibinfo{author}{\bibfnamefont{A.}~\bibnamefont{Stibor}},
  \bibinfo{journal}{New\ J.\ Phys.} \textbf{\bibinfo{volume}{22}},
  \bibinfo{pages}{063039} (\bibinfo{year}{2020}).

\bibitem[{\citenamefont{Chen and Batelaan}(2020)}]{CB20}
\bibinfo{author}{\bibfnamefont{Z.}~\bibnamefont{Chen}} \bibnamefont{and}
  \bibinfo{author}{\bibfnamefont{H.}~\bibnamefont{Batelaan}},
  \bibinfo{journal}{Europhys.\ Lett.} \textbf{\bibinfo{volume}{129}},
  \bibinfo{pages}{40004} (\bibinfo{year}{2020}).

\bibitem[{\citenamefont{Uhlemann et~al.}(2013)\citenamefont{Uhlemann, M\"uller,
  Hartel, Zach, and Haider}}]{UMH13}
\bibinfo{author}{\bibfnamefont{S.}~\bibnamefont{Uhlemann}},
  \bibinfo{author}{\bibfnamefont{H.}~\bibnamefont{M\"uller}},
  \bibinfo{author}{\bibfnamefont{P.}~\bibnamefont{Hartel}},
  \bibinfo{author}{\bibfnamefont{J.}~\bibnamefont{Zach}}, \bibnamefont{and}
  \bibinfo{author}{\bibfnamefont{M.}~\bibnamefont{Haider}},
  \bibinfo{journal}{Phys.\ Rev.\ Lett.} \textbf{\bibinfo{volume}{111}},
  \bibinfo{pages}{046101} (\bibinfo{year}{2013}).

\bibitem[{\citenamefont{Uhlemann et~al.}(2015)\citenamefont{Uhlemann, M\"uller,
  Zach, and Haider}}]{UMZ15}
\bibinfo{author}{\bibfnamefont{S.}~\bibnamefont{Uhlemann}},
  \bibinfo{author}{\bibfnamefont{H.}~\bibnamefont{M\"uller}},
  \bibinfo{author}{\bibfnamefont{J.}~\bibnamefont{Zach}}, \bibnamefont{and}
  \bibinfo{author}{\bibfnamefont{M.}~\bibnamefont{Haider}},
  \bibinfo{journal}{Ultramicroscopy} \textbf{\bibinfo{volume}{151}},
  \bibinfo{pages}{199} (\bibinfo{year}{2015}).

\bibitem[{\citenamefont{Martinetz et~al.}(2022)\citenamefont{Martinetz,
  Hornberger, and Stickler}}]{MHS22}
\bibinfo{author}{\bibfnamefont{L.}~\bibnamefont{Martinetz}},
  \bibinfo{author}{\bibfnamefont{K.}~\bibnamefont{Hornberger}},
  \bibnamefont{and} \bibinfo{author}{\bibfnamefont{B.~A.}
  \bibnamefont{Stickler}}, \bibinfo{journal}{Phys.\ Rev.\ X}
  \textbf{\bibinfo{volume}{3}}, \bibinfo{pages}{030327} (\bibinfo{year}{2022}).

\bibitem[{\citenamefont{Ford}(1993)}]{F93}
\bibinfo{author}{\bibfnamefont{L.~H.} \bibnamefont{Ford}},
  \bibinfo{journal}{Phys.\ Rev.\ D} \textbf{\bibinfo{volume}{47}},
  \bibinfo{pages}{5571} (\bibinfo{year}{1993}).

\bibitem[{\citenamefont{Ford}(1997)}]{F97}
\bibinfo{author}{\bibfnamefont{L.~H.} \bibnamefont{Ford}},
  \bibinfo{journal}{Phys.\ Rev.\ A} \textbf{\bibinfo{volume}{56}},
  \bibinfo{pages}{1812} (\bibinfo{year}{1997}).

\bibitem[{\citenamefont{Breuer and Petruccione}(2001)}]{BP01}
\bibinfo{author}{\bibfnamefont{H.~P.} \bibnamefont{Breuer}} \bibnamefont{and}
  \bibinfo{author}{\bibfnamefont{F.}~\bibnamefont{Petruccione}},
  \bibinfo{journal}{Phys.\ Rev.\ A} \textbf{\bibinfo{volume}{63}},
  \bibinfo{pages}{032102} (\bibinfo{year}{2001}).

\bibitem[{\citenamefont{Hsiang and Ford}(2004)}]{HF04}
\bibinfo{author}{\bibfnamefont{J.-T.} \bibnamefont{Hsiang}} \bibnamefont{and}
  \bibinfo{author}{\bibfnamefont{L.~H.} \bibnamefont{Ford}},
  \bibinfo{journal}{Phys.\ Rev.\ Lett.} \textbf{\bibinfo{volume}{92}},
  \bibinfo{pages}{250402} (\bibinfo{year}{2004}).

\bibitem[{\citenamefont{Alvarez and Mazzitelli}(2008)}]{AM08}
\bibinfo{author}{\bibfnamefont{E.}~\bibnamefont{Alvarez}} \bibnamefont{and}
  \bibinfo{author}{\bibfnamefont{F.}~\bibnamefont{Mazzitelli}},
  \bibinfo{journal}{Phys.\ Rev.\ A} \textbf{\bibinfo{volume}{77}},
  \bibinfo{pages}{032113} (\bibinfo{year}{2008}).

\bibitem[{\citenamefont{Hsiang and Ford}(2008)}]{HF08}
\bibinfo{author}{\bibfnamefont{J.-T.} \bibnamefont{Hsiang}} \bibnamefont{and}
  \bibinfo{author}{\bibfnamefont{L.~H.} \bibnamefont{Ford}},
  \bibinfo{journal}{Phys.\ Rev.\ D} \textbf{\bibinfo{volume}{78}},
  \bibinfo{pages}{065012} (\bibinfo{year}{2008}).

\bibitem[{\citenamefont{Zeilinger}(1999)}]{Z99_2}
\bibinfo{author}{\bibfnamefont{A.}~\bibnamefont{Zeilinger}},
  \bibinfo{journal}{Rev.\ Mod.\ Phys.} \textbf{\bibinfo{volume}{71}},
  \bibinfo{pages}{S288} (\bibinfo{year}{1999}).

\bibitem[{\citenamefont{Ma et~al.}(2016)\citenamefont{Ma, Kofler, and
  Zeilinger}}]{MKZ16}
\bibinfo{author}{\bibfnamefont{X.}~\bibnamefont{Ma}},
  \bibinfo{author}{\bibfnamefont{J.}~\bibnamefont{Kofler}}, \bibnamefont{and}
  \bibinfo{author}{\bibfnamefont{A.}~\bibnamefont{Zeilinger}},
  \bibinfo{journal}{Rev.\ Mod.\ Phys.} \textbf{\bibinfo{volume}{88}},
  \bibinfo{pages}{015005} (\bibinfo{year}{2016}).

\bibitem[{\citenamefont{Lichte and Freitag}(2000)}]{LF00}
\bibinfo{author}{\bibfnamefont{H.}~\bibnamefont{Lichte}} \bibnamefont{and}
  \bibinfo{author}{\bibfnamefont{B.}~\bibnamefont{Freitag}},
  \bibinfo{journal}{Ultramicroscopy} \textbf{\bibinfo{volume}{81}},
  \bibinfo{pages}{177} (\bibinfo{year}{2000}).

\bibitem[{\citenamefont{Herring}(2005)}]{H05}
\bibinfo{author}{\bibfnamefont{R.~A.} \bibnamefont{Herring}},
  \bibinfo{journal}{Ultramicroscopy} \textbf{\bibinfo{volume}{104}},
  \bibinfo{pages}{261} (\bibinfo{year}{2005}).

\bibitem[{\citenamefont{Verbeeck et~al.}(2008)\citenamefont{Verbeeck, Bertoni,
  and Schattschneider}}]{VBS08}
\bibinfo{author}{\bibfnamefont{J.}~\bibnamefont{Verbeeck}},
  \bibinfo{author}{\bibfnamefont{G.}~\bibnamefont{Bertoni}}, \bibnamefont{and}
  \bibinfo{author}{\bibfnamefont{P.}~\bibnamefont{Schattschneider}},
  \bibinfo{journal}{Ultramicroscopy} \textbf{\bibinfo{volume}{108}},
  \bibinfo{pages}{263} (\bibinfo{year}{2008}).

\bibitem[{\citenamefont{Schattschneider and L\"offler}(2018)}]{SL18}
\bibinfo{author}{\bibfnamefont{P.}~\bibnamefont{Schattschneider}}
  \bibnamefont{and}
  \bibinfo{author}{\bibfnamefont{S.}~\bibnamefont{L\"offler}},
  \bibinfo{journal}{Ultramicroscopy} \textbf{\bibinfo{volume}{190}},
  \bibinfo{pages}{39} (\bibinfo{year}{2018}).

\bibitem[{\citenamefont{Barwick et~al.}(2009)\citenamefont{Barwick, Flannigan,
  and Zewail}}]{BFZ09}
\bibinfo{author}{\bibfnamefont{B.}~\bibnamefont{Barwick}},
  \bibinfo{author}{\bibfnamefont{D.~J.} \bibnamefont{Flannigan}},
  \bibnamefont{and} \bibinfo{author}{\bibfnamefont{A.~H.}
  \bibnamefont{Zewail}}, \bibinfo{journal}{Nature}
  \textbf{\bibinfo{volume}{462}}, \bibinfo{pages}{902} (\bibinfo{year}{2009}).

\bibitem[{\citenamefont{{Garc\'{\i}a de Abajo}
  et~al.}(2010)\citenamefont{{Garc\'{\i}a de Abajo}, Asenjo-Garcia, and
  Kociak}}]{paper151}
\bibinfo{author}{\bibfnamefont{F.~J.} \bibnamefont{{Garc\'{\i}a de Abajo}}},
  \bibinfo{author}{\bibfnamefont{A.}~\bibnamefont{Asenjo-Garcia}},
  \bibnamefont{and} \bibinfo{author}{\bibfnamefont{M.}~\bibnamefont{Kociak}},
  \bibinfo{journal}{Nano\ Lett.} \textbf{\bibinfo{volume}{10}},
  \bibinfo{pages}{1859} (\bibinfo{year}{2010}).

\bibitem[{\citenamefont{Feist et~al.}(2015)\citenamefont{Feist, Echternkamp,
  Schauss, Yalunin, Sch\"afer, and Ropers}}]{FES15}
\bibinfo{author}{\bibfnamefont{A.}~\bibnamefont{Feist}},
  \bibinfo{author}{\bibfnamefont{K.~E.} \bibnamefont{Echternkamp}},
  \bibinfo{author}{\bibfnamefont{J.}~\bibnamefont{Schauss}},
  \bibinfo{author}{\bibfnamefont{S.~V.} \bibnamefont{Yalunin}},
  \bibinfo{author}{\bibfnamefont{S.}~\bibnamefont{Sch\"afer}},
  \bibnamefont{and} \bibinfo{author}{\bibfnamefont{C.}~\bibnamefont{Ropers}},
  \bibinfo{journal}{Nature} \textbf{\bibinfo{volume}{521}},
  \bibinfo{pages}{200} (\bibinfo{year}{2015}).

\bibitem[{\citenamefont{{Garc\'{\i}a de Abajo} and Ropers}(2023)}]{paper415}
\bibinfo{author}{\bibfnamefont{F.~J.} \bibnamefont{{Garc\'{\i}a de Abajo}}}
  \bibnamefont{and} \bibinfo{author}{\bibfnamefont{C.}~\bibnamefont{Ropers}},
  \bibinfo{journal}{Phys.\ Rev.\ Lett.} \textbf{\bibinfo{volume}{130}},
  \bibinfo{pages}{246901} (\bibinfo{year}{2023}).

\bibitem[{\citenamefont{{Di Giulio} and {Garc\'{\i}a de
  Abajo}}(2020)}]{paper357}
\bibinfo{author}{\bibfnamefont{V.}~\bibnamefont{{Di Giulio}}} \bibnamefont{and}
  \bibinfo{author}{\bibfnamefont{F.~J.} \bibnamefont{{Garc\'{\i}a de Abajo}}},
  \bibinfo{journal}{New\ J.\ Phys.} \textbf{\bibinfo{volume}{22}},
  \bibinfo{pages}{103057} (\bibinfo{year}{2020}).

\bibitem[{\citenamefont{{Garc\'{\i}a de Abajo}}(2009)}]{paper127}
\bibinfo{author}{\bibfnamefont{F.~J.} \bibnamefont{{Garc\'{\i}a de Abajo}}},
  \bibinfo{journal}{Phys.\ Rev.\ Lett.} \textbf{\bibinfo{volume}{102}},
  \bibinfo{pages}{237401} (\bibinfo{year}{2009}).

\bibitem[{\citenamefont{Lehmann and Lichte}(2002)}]{LL02}
\bibinfo{author}{\bibfnamefont{M.}~\bibnamefont{Lehmann}} \bibnamefont{and}
  \bibinfo{author}{\bibfnamefont{H.}~\bibnamefont{Lichte}},
  \bibinfo{journal}{Micros.\ Microanal.} \textbf{\bibinfo{volume}{8}},
  \bibinfo{pages}{447} (\bibinfo{year}{2002}).

\bibitem[{\citenamefont{Winkler et~al.}(2018)\citenamefont{Winkler, Barthel,
  Tavabi, Borghardt, Kardynal, and Dunin-Borkowski}}]{WBT18}
\bibinfo{author}{\bibfnamefont{F.}~\bibnamefont{Winkler}},
  \bibinfo{author}{\bibfnamefont{J.}~\bibnamefont{Barthel}},
  \bibinfo{author}{\bibfnamefont{A.~H.} \bibnamefont{Tavabi}},
  \bibinfo{author}{\bibfnamefont{S.}~\bibnamefont{Borghardt}},
  \bibinfo{author}{\bibfnamefont{B.~E.} \bibnamefont{Kardynal}},
  \bibnamefont{and} \bibinfo{author}{\bibfnamefont{R.~E.}
  \bibnamefont{Dunin-Borkowski}}, \bibinfo{journal}{Phys.\ Rev.\ Lett.}
  \textbf{\bibinfo{volume}{120}}, \bibinfo{pages}{156101}
  (\bibinfo{year}{2018}).

\bibitem[{\citenamefont{Horodecki et~al.}(2009)\citenamefont{Horodecki,
  Horodecki, Horodecki, and Horodecki}}]{HHH09}
\bibinfo{author}{\bibfnamefont{R.}~\bibnamefont{Horodecki}},
  \bibinfo{author}{\bibfnamefont{P.}~\bibnamefont{Horodecki}},
  \bibinfo{author}{\bibfnamefont{M.}~\bibnamefont{Horodecki}},
  \bibnamefont{and}
  \bibinfo{author}{\bibfnamefont{K.}~\bibnamefont{Horodecki}},
  \bibinfo{journal}{Rev.\ Mod.\ Phys.} \textbf{\bibinfo{volume}{81}},
  \bibinfo{pages}{865} (\bibinfo{year}{2009}).

\bibitem[{\citenamefont{Togan et~al.}(2010)\citenamefont{Togan, Chu, Trifonov,
  Jiang, Maze, Childress, Dutt, S\"{o}rensen, Hemmer, Zibrov et~al.}}]{TCT10}
\bibinfo{author}{\bibfnamefont{E.}~\bibnamefont{Togan}},
  \bibinfo{author}{\bibfnamefont{Y.}~\bibnamefont{Chu}},
  \bibinfo{author}{\bibfnamefont{A.~S.} \bibnamefont{Trifonov}},
  \bibinfo{author}{\bibfnamefont{L.}~\bibnamefont{Jiang}},
  \bibinfo{author}{\bibfnamefont{J.}~\bibnamefont{Maze}},
  \bibinfo{author}{\bibfnamefont{L.}~\bibnamefont{Childress}},
  \bibinfo{author}{\bibfnamefont{M.~V.~G.} \bibnamefont{Dutt}},
  \bibinfo{author}{\bibfnamefont{A.~S.} \bibnamefont{S\"{o}rensen}},
  \bibinfo{author}{\bibfnamefont{P.~R.} \bibnamefont{Hemmer}},
  \bibinfo{author}{\bibfnamefont{A.~S.} \bibnamefont{Zibrov}},
  \bibnamefont{et~al.}, \bibinfo{journal}{Nature}
  \textbf{\bibinfo{volume}{466}}, \bibinfo{pages}{730} (\bibinfo{year}{2010}).

\bibitem[{\citenamefont{Monroe et~al.}(1996)\citenamefont{Monroe, Meekhof,
  King, and Wineland}}]{MMK96}
\bibinfo{author}{\bibfnamefont{C.}~\bibnamefont{Monroe}},
  \bibinfo{author}{\bibfnamefont{D.~M.} \bibnamefont{Meekhof}},
  \bibinfo{author}{\bibfnamefont{B.~E.} \bibnamefont{King}}, \bibnamefont{and}
  \bibinfo{author}{\bibfnamefont{D.~J.} \bibnamefont{Wineland}},
  \bibinfo{journal}{Science} \textbf{\bibinfo{volume}{272}},
  \bibinfo{pages}{1131} (\bibinfo{year}{1996}).

\bibitem[{\citenamefont{Brune et~al.}(1996)\citenamefont{Brune, Hagley, Dreyer,
  Ma{\^i}tre, Maali, Wunderlich, Raimond, and Haroche}}]{BHD96}
\bibinfo{author}{\bibfnamefont{M.}~\bibnamefont{Brune}},
  \bibinfo{author}{\bibfnamefont{E.}~\bibnamefont{Hagley}},
  \bibinfo{author}{\bibfnamefont{J.}~\bibnamefont{Dreyer}},
  \bibinfo{author}{\bibfnamefont{X.}~\bibnamefont{Ma{\^i}tre}},
  \bibinfo{author}{\bibfnamefont{A.}~\bibnamefont{Maali}},
  \bibinfo{author}{\bibfnamefont{C.}~\bibnamefont{Wunderlich}},
  \bibinfo{author}{\bibfnamefont{J.}~\bibnamefont{Raimond}}, \bibnamefont{and}
  \bibinfo{author}{\bibfnamefont{S.}~\bibnamefont{Haroche}},
  \bibinfo{journal}{Phys.\ Rev.\ Lett.} \textbf{\bibinfo{volume}{77}},
  \bibinfo{pages}{4887} (\bibinfo{year}{1996}).

\bibitem[{\citenamefont{{Garc\'{\i}a de Abajo}}(2013)}]{paper228}
\bibinfo{author}{\bibfnamefont{F.~J.} \bibnamefont{{Garc\'{\i}a de Abajo}}},
  \bibinfo{journal}{ACS\ Nano} \textbf{\bibinfo{volume}{7}},
  \bibinfo{pages}{11409} (\bibinfo{year}{2013}).

\bibitem[{\citenamefont{Gradshteyn and Ryzhik}(2007)}]{GR1980}
\bibinfo{author}{\bibfnamefont{I.~S.} \bibnamefont{Gradshteyn}}
  \bibnamefont{and} \bibinfo{author}{\bibfnamefont{I.~M.}
  \bibnamefont{Ryzhik}}, \emph{\bibinfo{title}{Table of Integrals, Series, and
  Products}} (\bibinfo{publisher}{Academic Press}, \bibinfo{address}{London},
  \bibinfo{year}{2007}).

\bibitem[{\citenamefont{Johnson et~al.}(2021)\citenamefont{Johnson, Turner, and
  McMorran}}]{JTM21}
\bibinfo{author}{\bibfnamefont{C.~W.} \bibnamefont{Johnson}},
  \bibinfo{author}{\bibfnamefont{A.~E.} \bibnamefont{Turner}},
  \bibnamefont{and} \bibinfo{author}{\bibfnamefont{B.~J.}
  \bibnamefont{McMorran}}, \bibinfo{journal}{Phys.\ Rev.\ Research}
  \textbf{\bibinfo{volume}{3}}, \bibinfo{pages}{043009} (\bibinfo{year}{2021}).

\bibitem[{\citenamefont{Johnson et~al.}(2022)\citenamefont{Johnson, Turner,
  {Garc\'{\i}a de Abajo}, and McMorran}}]{paper388}
\bibinfo{author}{\bibfnamefont{C.~W.} \bibnamefont{Johnson}},
  \bibinfo{author}{\bibfnamefont{A.~E.} \bibnamefont{Turner}},
  \bibinfo{author}{\bibfnamefont{F.~J.} \bibnamefont{{Garc\'{\i}a de Abajo}}},
  \bibnamefont{and} \bibinfo{author}{\bibfnamefont{B.~J.}
  \bibnamefont{McMorran}}, \bibinfo{journal}{Phys.\ Rev.\ Lett.}
  \textbf{\bibinfo{volume}{128}}, \bibinfo{pages}{147401}
  (\bibinfo{year}{2022}).

\bibitem[{\citenamefont{{Di Giulio} et~al.}(2019)\citenamefont{{Di Giulio},
  Kociak, and {Garc\'{\i}a de Abajo}}}]{paper339}
\bibinfo{author}{\bibfnamefont{V.}~\bibnamefont{{Di Giulio}}},
  \bibinfo{author}{\bibfnamefont{M.}~\bibnamefont{Kociak}}, \bibnamefont{and}
  \bibinfo{author}{\bibfnamefont{F.~J.} \bibnamefont{{Garc\'{\i}a de Abajo}}},
  \bibinfo{journal}{Optica} \textbf{\bibinfo{volume}{6}}, \bibinfo{pages}{1524}
  (\bibinfo{year}{2019}).

\bibitem[{not({\natexlab{a}})}]{notedecoherence1}
\bibinfo{note}{{This identity is obtained by performing a Taylor expansion of
  the exponential in $\ee^{\vb t\cdot\nabla}f(\rb)=\sum_{n=0}^\infty[(\vb
  t\cdot\nabla)^n/n!]f(\rb)$, such that the right-hand side coincides with the
  Taylor expansion of $f(\rb+\vb t)$ around $\rb$. From here, we ready verify
  the equation $f(\rb)\ee^{-\vb t\cdot\nabla}g(\rb)=\ee^{-\vb
  t\cdot\nabla}f(\rb+\vb t)g(\rb)$ for arbitrary functions $f(\rb)$ and
  $g(\rb)$.}}

\bibitem[{\citenamefont{Magnus}(1954)}]{M1954}
\bibinfo{author}{\bibfnamefont{W.}~\bibnamefont{Magnus}},
  \bibinfo{journal}{Comm.\ Pure\ Appl.\ Math.} \textbf{\bibinfo{volume}{VII}},
  \bibinfo{pages}{649} (\bibinfo{year}{1954}).

\bibitem[{not({\natexlab{b}})}]{notedecoherence2}
\bibinfo{note}{{Dropping the $j$ subindex for simplicity, we first write
  $\ee^{\beta^*\aa^\dagger-\beta\aa}=\ee^{-|\beta|^2/2}\ee^{\beta^*\aa^\dagger}\ee^{-\beta\aa}$
  as a direct consequence of the Baker-Campbell-Hausdorff formula. We now apply
  the thermal average defined by $\langle\hat{C}\rangle_T=\sum_{n=0}^\infty
  \bra{n}\hat{C}\ket{n}\,p_n$ for any operator $\hat{C}$ in terms of the
  occupation numbers $p_n=(1-\ee^{-\theta})\ee^{-n\theta}$ with
  $\theta=\hbar\omega/\kB T$, considering a mode of frequency $\omega$ at
  temperature $T$. By Taylor-expanding the above exponentials of operators, and
  noticing that $\bra{n}(\aa^\dagger)^j\aa^{j'}\ket{n}=\delta_{jj'}n!/(n-j)!$
  with $j\le n$ (i.e., only $j=j'$ terms survive), we find
  $\langle\ee^{\beta^*\aa^\dagger-\beta\aa}\rangle_T=\ee^{-|\beta|^2/2}\sum_{j=0}^\infty\big[(-|\beta|^2)^j/j!\big]\,S_j$
  with $S_j=\sum_{n=0}^\infty p_n \binom{n}{j}$. Then, using the relation$^3$
  $S_j=\bar{n}^j$, where $\bar{n}=1/(\ee^\theta-1)$ denotes the average
  population (i.e., the Bose-Einstein distribution), we obtain
  $\big\langle\ee^{\beta^*\aa^\dagger-\beta\aa}\big\rangle_T=\ee^{-(\bar{n}+1/2)\,|\beta|^2}$.
  Finally, this expression becomes
  $\big\langle\ee^{\beta^*\aa^\dagger-\beta\aa}\big\rangle_T=\exp\big\{(1/2)\big\langle(\beta^*\aa^\dagger-\beta\aa)^2\big\rangle_T\big\}$
  by comparing the exponent to the thermal average of
  $(\beta^*\aa^\dagger-\beta\aa)^2$.}}

\bibitem[{not({\natexlab{c}})}]{notedecoherence3}
\bibinfo{note}{{The sum $S_j=\sum_{n=0}^\infty p_n
  \binom{n}{j}=[(1-\ee^{-\theta})/j!]\sum_{n=0}^\infty
  n(n-1)\cdots(n-j+1)\,\ee^{-n\theta}$ is directly found to satisfy the
  recursion relation $S_{j+1}=(\bar{n}-j-\partial_\theta)\,S_j/(j+1)$, which
  trivially admits the solution $S_j=\bar{n}^j$ with
  $\bar{n}=1/(\ee^\theta-1)$.}}

\bibitem[{\citenamefont{{Garc\'{\i}a de Abajo} and {Di
  Giulio}}(2021)}]{paper371}
\bibinfo{author}{\bibfnamefont{F.~J.} \bibnamefont{{Garc\'{\i}a de Abajo}}}
  \bibnamefont{and} \bibinfo{author}{\bibfnamefont{V.}~\bibnamefont{{Di
  Giulio}}}, \bibinfo{journal}{ACS\ Photonics} \textbf{\bibinfo{volume}{8}},
  \bibinfo{pages}{945} (\bibinfo{year}{2021}).

\bibitem[{\citenamefont{Buhmann}(2012)}]{B12_2}
\bibinfo{author}{\bibfnamefont{S.~Y.} \bibnamefont{Buhmann}},
  \emph{\bibinfo{title}{Dispersion {F}orces {I}. {M}acroscopic {Q}uantum
  {E}lectrodynamics and {G}round-{S}tate {C}asimir, {C}asimir-{P}older and
  {van} {der} {Waals} {F}orces}} (\bibinfo{publisher}{Springer-Verlag Berlin
  Heidelberg}, \bibinfo{address}{Verlag Berlin Heidelberg},
  \bibinfo{year}{2012}).

\bibitem[{\citenamefont{{Garc\'{\i}a de Abajo}}(2010)}]{paper149}
\bibinfo{author}{\bibfnamefont{F.~J.} \bibnamefont{{Garc\'{\i}a de Abajo}}},
  \bibinfo{journal}{Rev.\ Mod.\ Phys.} \textbf{\bibinfo{volume}{82}},
  \bibinfo{pages}{209} (\bibinfo{year}{2010}).

\bibitem[{\citenamefont{Jackson}(1999)}]{J99}
\bibinfo{author}{\bibfnamefont{J.~D.} \bibnamefont{Jackson}},
  \emph{\bibinfo{title}{Classical Electrodynamics}}
  (\bibinfo{publisher}{Wiley}, \bibinfo{address}{New York},
  \bibinfo{year}{1999}).

\bibitem[{\citenamefont{Born and Wolf}(1999)}]{BW99}
\bibinfo{author}{\bibfnamefont{M.}~\bibnamefont{Born}} \bibnamefont{and}
  \bibinfo{author}{\bibfnamefont{E.}~\bibnamefont{Wolf}},
  \emph{\bibinfo{title}{Principles of Optics: Electromagnetic Theory of
  Propagation, Interference and Diffraction of Light}}
  (\bibinfo{publisher}{Cambridge University Press},
  \bibinfo{address}{Cambridge}, \bibinfo{year}{1999}).

\bibitem[{\citenamefont{Jones}(1950)}]{J1950}
\bibinfo{author}{\bibfnamefont{D.~S.} \bibnamefont{Jones}},
  \bibinfo{journal}{Quart.\ J.\ Mech.\ Appl.\ Math.}
  \textbf{\bibinfo{volume}{3}}, \bibinfo{pages}{420} (\bibinfo{year}{1950}).

\bibitem[{not({\natexlab{d}})}]{notedecoherence4}
\bibinfo{note}{{In Eq.~(\ref{jconds1}), $k_z$ introduces two branch cuts at
  $k_x=\pm K$ [see Eq.~(\ref{kzkk})], but the one at $k_x=-K$ is canceled by
  the numerator of Eq.~(\ref{jksol}).}}

\bibitem[{\citenamefont{{van de Hulst}}(1981)}]{V1981}
\bibinfo{author}{\bibfnamefont{H.~C.} \bibnamefont{{van de Hulst}}},
  \emph{\bibinfo{title}{Light Scattering by Small Particles}}
  (\bibinfo{publisher}{Dover}, \bibinfo{address}{New York},
  \bibinfo{year}{1981}).

\bibitem[{\citenamefont{Nyquist}(1928)}]{N1928}
\bibinfo{author}{\bibfnamefont{H.}~\bibnamefont{Nyquist}},
  \bibinfo{journal}{Phys.\ Rev.} \textbf{\bibinfo{volume}{32}},
  \bibinfo{pages}{110} (\bibinfo{year}{1928}).

\bibitem[{\citenamefont{Callen and Welton}(1951)}]{CW1951}
\bibinfo{author}{\bibfnamefont{H.~B.} \bibnamefont{Callen}} \bibnamefont{and}
  \bibinfo{author}{\bibfnamefont{T.~A.} \bibnamefont{Welton}},
  \bibinfo{journal}{Phys.\ Rev.} \textbf{\bibinfo{volume}{83}},
  \bibinfo{pages}{34} (\bibinfo{year}{1951}).

\bibitem[{\citenamefont{Abrikosov et~al.}(1965)\citenamefont{Abrikosov, Gorkov,
  and Dzyaloshinskii}}]{AGD1965}
\bibinfo{author}{\bibfnamefont{A.~A.} \bibnamefont{Abrikosov}},
  \bibinfo{author}{\bibfnamefont{L.~P.} \bibnamefont{Gorkov}},
  \bibnamefont{and} \bibinfo{author}{\bibfnamefont{I.~Y.}
  \bibnamefont{Dzyaloshinskii}}, \emph{\bibinfo{title}{Quantum Field
  Theoretical Methods in Statistical Physics}} (\bibinfo{publisher}{Pergamon
  Press}, \bibinfo{address}{New York}, \bibinfo{year}{1965}).

\bibitem[{\citenamefont{Bonch-Bruevich and Tyablikov}(1962)}]{BT1962}
\bibinfo{author}{\bibfnamefont{V.~L.} \bibnamefont{Bonch-Bruevich}}
  \bibnamefont{and} \bibinfo{author}{\bibfnamefont{S.~V.}
  \bibnamefont{Tyablikov}}, \emph{\bibinfo{title}{The Green Function Method in
  Statistical Mechanics}} (\bibinfo{publisher}{North Holland},
  \bibinfo{address}{Amsterdam}, \bibinfo{year}{1962}).

\bibitem[{not({\natexlab{e}})}]{notedecoherence5}
\bibinfo{note}{{The functions $\sin(\omega|\tau|)$ and $\Imm\{\chi(\omega)\}$
  are both odd in $\omega$, so we can write the integral in the left-hand side
  of Eq.~(\ref{sfsc}) as $(1/2)\int_{-\infty}^\infty
  d\omega\;\sin(\omega|\tau|)\;\Imm\{\chi(\omega)\}$. In addition, by Fourier
  transforming the sine function, the inverse Fourier transform yields
  $\sin(\omega|\tau|)=(1/2\pi){\rm P}\!\int_{-\infty}^\infty\!\! d\omega'
  [1/(\omega+\omega')+1/(\omega-\omega')]\,\ee^{-\ii\omega'\tau}$, where P
  stands for the principal value. After making this substitution in
  Eq.~(\ref{sfsc}), the $\omega$ integral can directly be performed by using
  the Kramers-Kronig relation ${\rm P}\!\int_{-\infty}^\infty\!\!
  d\omega\;\Imm\{\chi(\omega)\}/(\omega\pm\omega')=\pi\,{\rm
  Re}\{\chi(\omega')\}$. Finally, noticing the parity of the remaining
  functions in the integrand and changing $\omega'$ to $\omega$, we obtain
  Eq.~(\ref{sfsc}).}}

\end{thebibliography}


\clearpage 
\pagebreak \onecolumngrid \section*{SUPPLEMENTARY FIGURES}
\renewcommand{\thefigure}{S\arabic{figure}}
\setcounter{figure}{0}

\begin{figure*}[th!]
\centering{\includegraphics[width=0.62\textwidth]{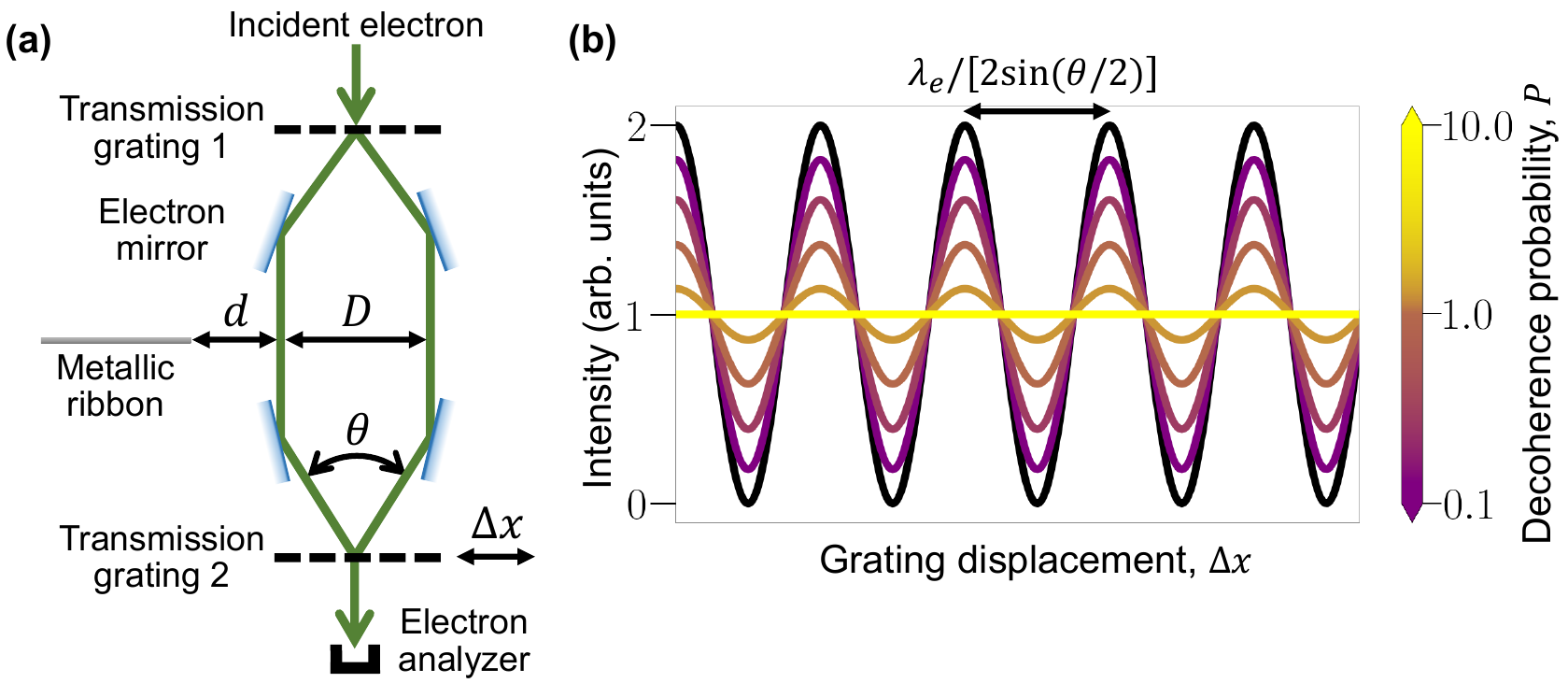}}
\caption{\textbf{A possible configuration for measuring electron decoherence with large inter-path separations.} (a)~Electron optics scheme incorporating a transmission grating that splits the incident beam into two paths; two upper electrostatic mirrors that redirect the paths such they run parallel to each other with a separation $D$ at the position of the specimen (e.g., a wide metallic ribbon); two subsequent mirrors that make the paths converge; a second movable transmission grating that mixes the two paths into a single Bragg transmission direction; and an electron detector. The second grating must have a period $\lambda_e/\sin(\theta/2)$ determined by the angle $\theta$ between the two convergent paths and the electron wavelength $\lambda_e$, such that Bragg scattering combines both of them along a direction normal to the grating. The electron signal oscillates with the lateral displacement of the grating $\Delta x$, provided the two paths maintain some degree of coherence. (b)~Interference pattern as a function of grating 2 displacement $\Delta x$ for different values of the decoherence probability produced by radiative coupling mediated by a metallic half-plane under the configuration of (a) (see specific examples in Figs.~1 and \ref{FigS2} below). The interference patterns have a period $\lambda_e/[2\sin(\theta/2)]$ in $\Delta x$. Gratings similar to those used in previous studies \cite{JTM21,paper388} would fit in the current scheme.}
\label{FigS1}
\end{figure*}

\begin{figure*}[th!]
\centering{\includegraphics[width=0.75\textwidth]{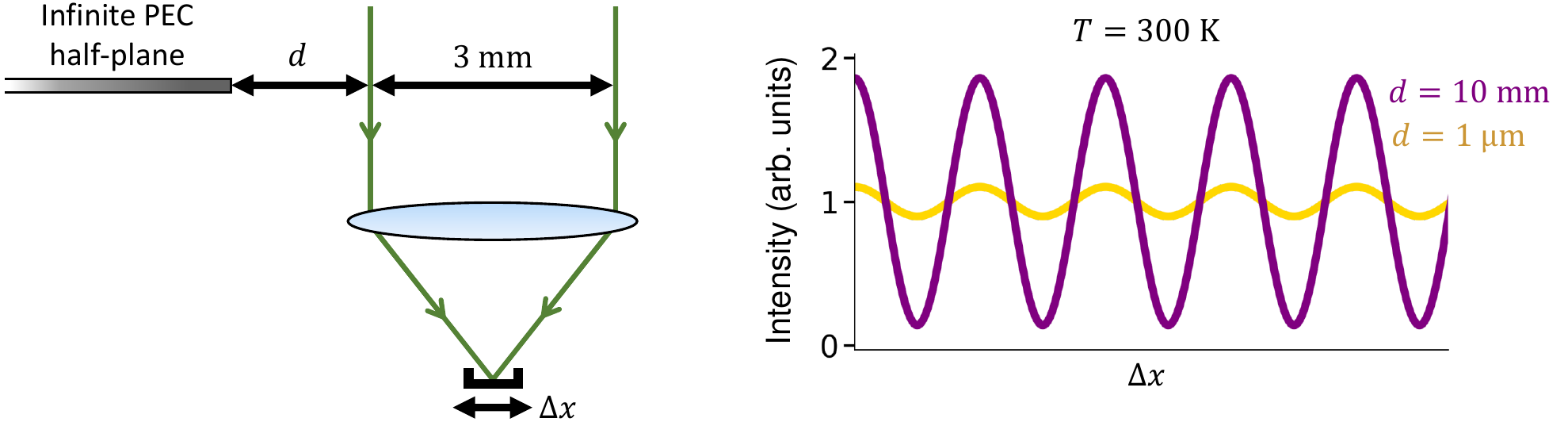}}
\caption{{\bf Depletion of fringe visibility in a feasible experimental scenario.} We consider a configuration similar to Fig.~1c in the main text, where a two-path electron interacts with a perfect-electric-conductor (PEC) half-plane at a temperature of 300~K and undergoes a degree of decoherence that is observed through the visibility of the interference fringes obtained as a function of transverse position $\Delta x$ (or deflection angle) when mixing both paths at a detector (see also Fig.~\ref{FigS1}). Here, we fix the inter-path separation to 3~mm and consider two different electron-edge distances: $d=1$~$\mu$m and 10~$\mu$m. A substantial reduction in fringe visibility is predicted for the shortest $d$.}
\label{FigS2}
\end{figure*}

\begin{figure*}[th!]
\centering{\includegraphics[width=0.83\textwidth]{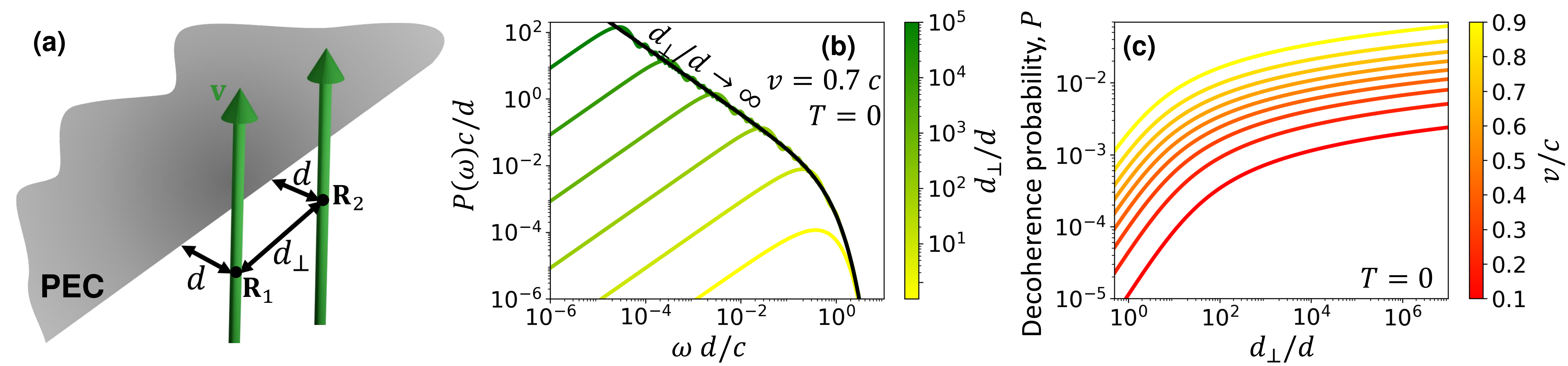}}
\caption{{\bf Two-path electron decoherence by a half-plane at zero temperature in a laterally displaced arrangement.} (a)~System under consideration, consisting of a single electron split into a two-path spatial superposition and passing close and perpendicularly to a perfectly conducting half-plane at a distance $d$ from the edge with a lateral separation $d_\perp$ [beam path positions $\Rb_1=(d,0)$ and $\Rb_2=(d,d_\perp)$]. (b)~Universal plot of the spectrally resolved decoherence probability for various $d_\perp/d$ ratios (color curves), approaching a divergent profile in the $d_\perp\gg d$ limit (black curve). We consider an electron velocity $v=0.7\,c$ and normalize the frequency and the probability using the distance $d$. (c)~Decoherence probability as a function of $d_\perp/d$ for different electron velocities (see color scale).}
\label{FigS3}
\end{figure*}

\begin{figure*}[th!]
\centering{\includegraphics[width=0.7\textwidth]{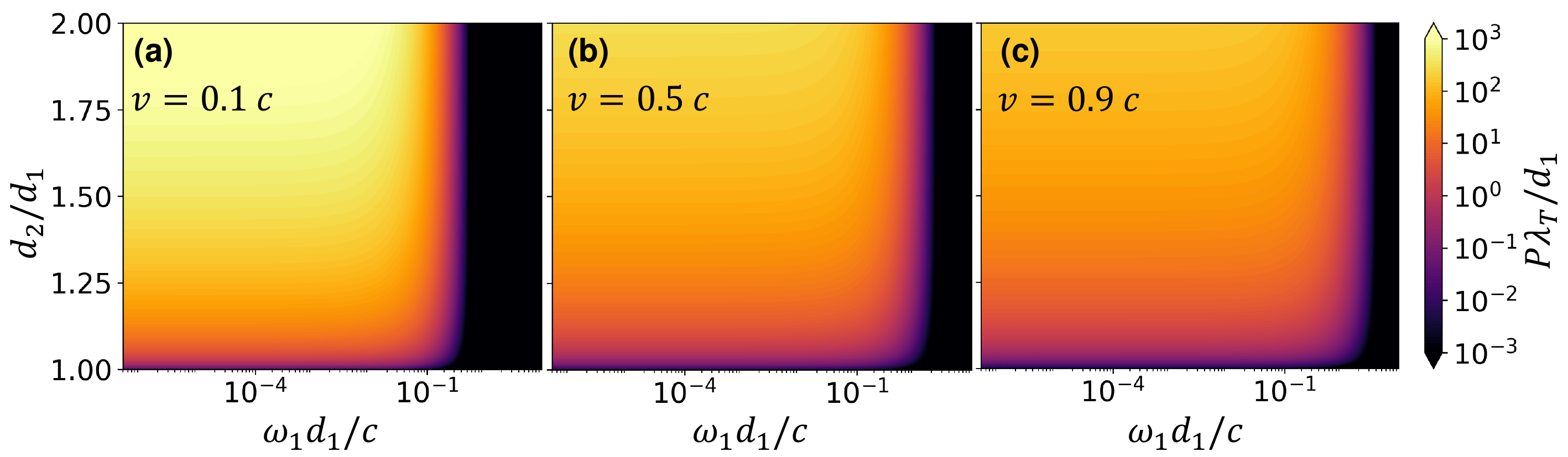}}
\caption{{\bf Spectral decomposition of the decoherence probability by a half-plane at finite temperature.} We consider the configuration of Fig.~2a in the main text with $d_\perp=0$ and present results for three different electron velocities in panels (a)-(c) (see labels).}
\label{FigS4}
\end{figure*}

\begin{figure*}[th!]
\centering{\includegraphics[width=0.83\textwidth]{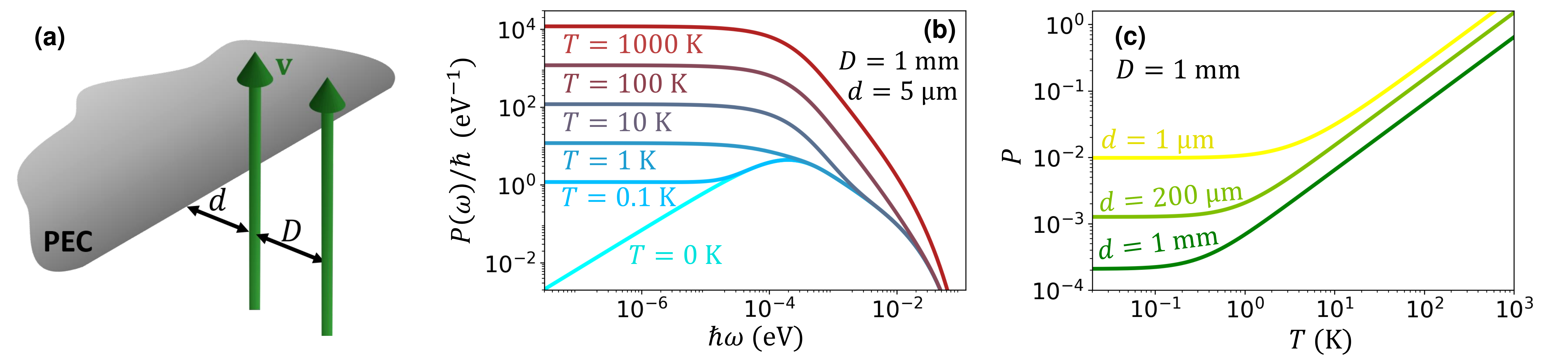}}
\caption{\textbf{Onset of temperature effects in electron decoherence.} (a)~We consider a two-path electron beam passing near a metallic half-plane with electron-edge and inter-path separations $d$ and $D$, respectively. (b)~Spectral decomposition of the decoherence probability for $d=5$~$\mu$m, $D=500$~$\mu$m, and different temperatures in the $T=0-1000$~K range. (c)~Temperature dependence of the decoherence probability for $D=500$~$\mu$m and various values of $d$.}
\label{FigS5}
\end{figure*}

\begin{figure*}[th!]
\centering{\includegraphics[width=0.5\textwidth]{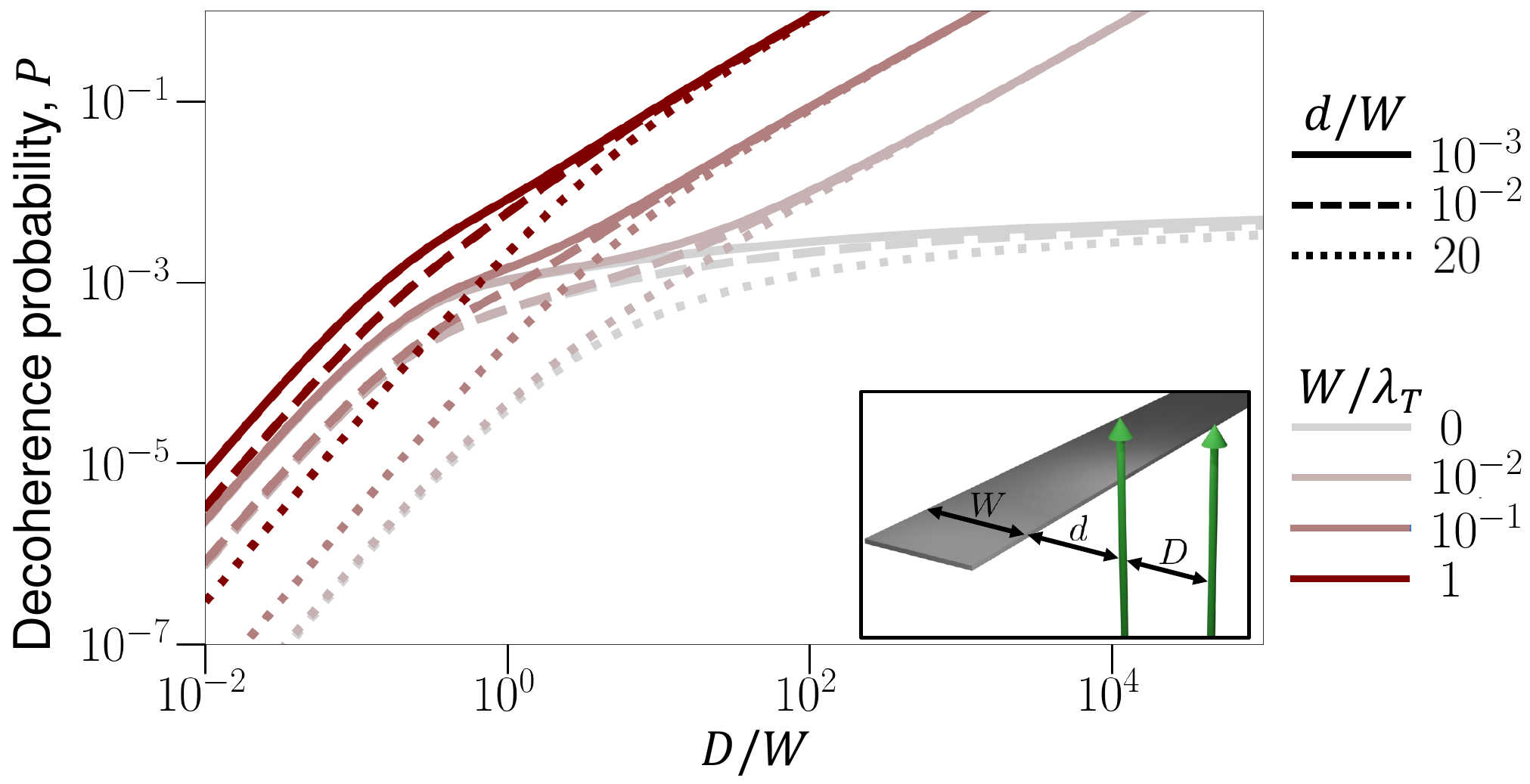}}
\caption{{\bf Finite size effects on the decoherence probability.} Similar to Fig.~4 in the main text, but with the decoherence probability $P$ plotted as a function of the inter-path separation $D$ normalized to the ribbon width $W$. The electron beam is prepared in a two-path superposition state (inter-path distance $D$) and interacts with a ribbon of width $W$. The path--ribbon distances are $d$ and $d+D$. We present results for different values of the $d/W$ and $W/\lambda_T$ ratios (see the legend on the right).}
\label{FigS6}
\end{figure*}

\end{document}